\documentclass[10pt]{iopart}

\usepackage{graphicx}
\usepackage{amssymb}
\usepackage{exscale}
\usepackage{iopams}
\usepackage{xspace}
\usepackage{citesort}
\usepackage{times}

\newfont{\tensy}{cmsy10}
\newcommand{\chem}[1]{{$\fontdimen16\tensy=3.0pt
    \fontdimen17\tensy=3.0pt \mathrm{#1}$}}

\renewcommand{\Im}[0]{\mathrm{Im}}
\renewcommand{\Re}[0]{\mathrm{Re}}
\newcommand{\ie}[0]{i.e.\@\xspace}
\newcommand{\eg}[0]{e.g.\@\xspace}
\newcommand{\om}[0]{\omega}
\newcommand{\en}[0]{\epsilon}
\newcommand{\gammab}[0]{\overline{\gamma}}
\newcommand{\Ep}{E_\mathrm{P}}
\newcommand{\Ap}[0]{A_\mathrm{p}}
\newcommand{\Ae}[0]{A_\mathrm{e}}
\newcommand{\nag}{\phantom{\dag}}
\newcommand{\Ek}{E_k}
\newcommand{\Sr}{S^\mathrm{reg}}
\newcommand{\sir}{\sigma^\mathrm{reg}}
\newcommand{\D}{\mathcal{D}}
\newcommand{\las}[0]{\langle}
\newcommand{\ras}[0]{\rangle}
\renewcommand{\hat}[1]{\widehat{#1}}
\renewcommand{\tilde}[1]{\widetilde{#1}}

\begin{document}

\renewcommand\floatpagefraction{.9}
\renewcommand\topfraction{.9}
\renewcommand\bottomfraction{.9}
\renewcommand\textfraction{.1}   

\title{Optical conductivity of polaronic charge carriers}

\author{J Loos\dag, M Hohenadler\ddag, A Alvermann\S, and H Fehske\S}

\address{\dag\ %
  Institute of Physics, Academy of Sciences of the Czech Republic, Prague
}

\address{\ddag\ %
  Theory of Condensed Matter, University of Cambridge, United Kingdom
}

\address{\S\ %
  Institute of Physics, Ernst-Moritz-Arndt University Greifswald, Germany
}

\ead{\mailto{loos@fzu.cz}}

\begin{abstract}
  The optical conductivity of charge carriers coupled to quantum phonons is studied
  in the framework of the one-dimensional spinless Holstein model. For one
  electron, variational diagonalisation yields exact results in the
  thermodynamic limit, whereas at finite carrier density analytical
  approximations based on previous work on single-particle spectral functions
  are obtained. Particular emphasis is put on deviations from weak-coupling,
  small-polaron or one-electron theories occurring at intermediate coupling
  and/or finite carrier density. The analytical results are in surprisingly
  good agreement with exact data, and exhibit the characteristic
  polaronic excitations observed in experiments on manganites.
\end{abstract}

\pacs{78.20.Bh, 71.10.-w, 71.27.+a, 75.47.Lx, 78.20.-e}


\section{Introduction}\label{sec:introduction}

In recent years, there has been a lot of controversy about the possible
interpretation of optical data on strongly correlated electron systems such
as manganites, cuprates and nickelates in terms of polaronic charge carriers
\cite{CaDoLuPaMaGiRuChSa97,SaAlLi95}.  It turns
out that simple one-electron or perturbative weak/strong-coupling theories
can usually not explain more than one type of experiment for a given set of
(realistic) parameters, and a general discussion of
this issue for the case of the manganites has been given in
\cite{HoEd01,David_AiP,HaMaLoKo04}. Owing to these discrepancies, and also in the
light of recent progress in the understanding of many-polaron systems
\cite{CaGrSt99,HoNevdLWeLoFe04,HoHaWeFe06}, it is highly desirable to revisit this
problem using both analytical and numerical many-body methods.

Ignoring the necessity of taking into account cooperative Jahn-Teller effects
to describe orbital effects in manganites \cite{David_AiP}, an important
issue in the framework of one-band models concerns the validity of two
complementary types of electron-phonon interaction, namely the Holstein
\cite{Ho59a} and the Fr\"{o}hlich model \cite{Fr54} for screened, local
coupling and unscreened, long-range electron-phonon interaction,
respectively. Motivated by recent experimental results and the absence of a
unified theory for, \eg, ferromagnetic manganites \cite{HaMaLoKo04}, we
analyse the signatures of polaronic excitations in the optical conductivity,
both in the low-density limit and at finite carrier density, in the framework
of the one-dimensional spinless Holstein model.

Polaronic materials such as the manganites generally require a theory valid
for finite carrier density and all coupling regimes
\cite{David_AiP,HaMaLoKo04}. Whereas a large number of works have reported on
approximate findings for the one-electron limit (the polaron problem) of the
Holstein model, here we present variational diagonalisation results for the
thermodynamic limit, thereby providing an exact (numerical) solution. Exact
data for finite clusters, also at finite temperature, can be found in
\cite{ScWeWeAlFe05}.

The many-electron case has been studied by approximate methods in the past
\cite{PhysRevB.44.7127,FeIhLoTrBu94,PeFiCaIa01,HoEd01}. Moreover, exact results
on finite clusters are available
\cite{FeWeHaWeBi03,WeBiHoScFe05,WeBiHoScFe05}. In order to further improve
the present understanding, we examine the optical conductivity in the
framework of electronic spectral functions deduced by perturbative and
variational methods valid at finite carrier density
\cite{LoHoFe06,LoHoAlFe06}. The important advantage of such calculations is
the possibility to relate the observed contributions to the optical response
to specific processes (transitions) in the analytical formulas, and to
understand the relevance of these contributions in dependence on the model
parameters. Special attention will be paid to the role played by the coherent
(respectively incoherent) parts of the spectral functions.  We compare the
analytical results to exact numerical data, and discuss the differences
between the weak-coupling (WC), intermediate-coupling (IC) and
strong-coupling (SC) regimes, as well as the influence of carrier density and
phonon frequency. Besides, we attempt to make a qualitative connection of our
findings to recent experiments on manganites \cite{HaMaLoKo04}, as well as to
theoretical work on the many-electron case based on the Holstein model
\cite{HoEd01,WeBiHoScFe05} and the Fr\"{o}hlich model
\cite{TeDe01,GuLaFi62}.  Theoretical work on manganites is discussed in
\cite{HoEd01,David_AiP,HaMaLoKo04}, and a review of polaron theories can be
found in \cite{AlMo95}.

This paper is organised as follows. In section~\ref{sec:model}, we introduce
the spinless Holstein model. In section~\ref{sec:theory}, we outline the
derivation of the general expression for the optical conductivity involving
the electronic spectral functions deduced in \cite{LoHoFe06}.  Numerical and
analytical results are discussed in section~\ref{sec:results}, and
section~\ref{sec:conclusion} contains our conclusions.

\section{Model}\label{sec:model}
 
We consider the one-dimensional (1D) spinless Holstein model, describing
fermions coupled to dispersionless optical phonons. It provides a general
framework to study polaron physics \cite{FeAlHoWe06}, many-polaron effects
\cite{HoHaWeFe06} and quantum phase transitions
\cite{SyHuBeWeFe04,CrSaCa05,HoWeBiAlFe06,SyHuBe05}, but is simple enough to
permit reliable investigations by analytical and numerical methods. A
connection to the more general Holstein-Hubbard model (see, \eg,
\cite{FeWeHaWeBi03}) can be made in the limit of large Hubbard-$U$ in the
latter. Moreover, the Holstein double-exchange model for the manganites
\cite{David_AiP,Gr01,HoEd01} reduces to the present model for infinite Hund's
rule coupling and a ferromagnetic state.

Following \cite{LoHoFe06}, we write the Hamiltonian in the general form
\begin{equation}\label{eq:HM}
  H 
  =
    \eta\sum_i c^\dag_i c^{\nag}_i 
  - \sum_{i,j}C_{ij}^{\nag} c_{i}^{\dag} c_{j}^{\nag}
  + \om_0\sum_i  
  ( b_i^{\dag} b_i^{\nag} + \mbox{\small $\frac{1}{2}$})
  \,,
\end{equation}
where $c_{i}^{\dag}$ ($c_{i}^{\nag}$) and $b_i^{\dag}$ ($b_i^{\nag}$) create
(annihilate) a spinless fermion respectively a phonon of energy $\om_0$
($\hbar=1$) at site $i$. The strength of the electron-phonon interaction is
specified by the dimensionless coupling constants $\lambda=\Ep/2t$ (adiabatic
regime, $\om_0/t\ll1$) and $g^2=\Ep/\om_0$ (anti-adiabatic regime,
$\om_0/t\gg1$), with $\Ep$ denoting the atomic-limit [$C_{ij}=0$ for $i\neq
j$ in equation~(\ref{eq:HM})] polaron binding energy.

The definitions of $\eta$ and the coefficients $C_{ij}$ will depend on the
type of approximation used and hence on the parameter regime. In the WC case,
in which we start with the original, untransformed Holstein Hamiltonian, we
have
\begin{equation} \label{eq:wcC}
  \eta=-\mu
  \,,\quad
  C_{ii} 
  =
  g \om_0  ( b_i^{\dag} + b_i^{\nag}) 
  \,,\quad 
  C_{\las ij\ras}
  =
  t
  \,,
\end{equation}
where $\mu$ denotes the chemical potential. In contrast, in the SC regime, we
use the Hamiltonian after the Lang-Firsov transformation \cite{LangFirsov} with
\begin{equation}\label{eq:scC}
  \eta=-\Ep-\mu
  \,,\quad
  C_{ii}
  =
  0 
  \,,\quad 
  C_{\las ij\ras}
  =
  t \rme^{-g( b_i^{\dag} - b_i^{\nag} - b_j^{\dag} + b_j^{\nag})} 
  \,.
\end{equation}
The non-interacting (polaron) half-bandwidth in the SC regime is
$\tilde{W}=W\rme^{-g^2}$, with the 1D free-fermion half-bandwidth
$W=2t$. Finally, the treatment of the IC case is based on a modified Lang-Firsov
transformation \cite{LoHoFe06} defined by the unitary operator
\begin{equation}\label{eq:ic_transformation}
  U 
  =
  \rme^{\sum_i g (\gammab c^\dag_i c^{\nag}_i + \gamma) (b^\dag_i -
    b^{\nag}_i)}
  \,,
\end{equation}
which leads to a Hamiltonian of the form~(\ref{eq:HM}) with
\begin{eqnarray}\label{eq:trans_ic}\nonumber
  \eta&=&-\mu -\Ep [\gammab (2-\gammab) + 2\gamma (1-\gammab)]
  \,,
  \\
  C_{ii}
  &=&
  g\omega_0 (1-\gammab)(b^\dag_i+b^{\nag}_i)
  \,,\quad
  C_{\las ij \ras}
  =
  t \rme^{-\gammab g (b^\dag_i - b^{\nag}_i - b^\dag_j + b^{\nag}_j)}
  \,.
\end{eqnarray}
The parameters $\gamma$ and $\gammab$ depend on the carrier concentration
$n$, and on the variational parameter $R$ characterising the delocalisation
of the charge carrier from the center of the associated lattice
distortion. Explicitly, we have
\begin{equation}\label{eq:gammab}
  \gamma = 2 n\rme^{-1/R} \tanh\frac{1}{2R}
  \,,\quad
  \gammab =  \tanh\frac{1}{2R}-\gamma
  \,,
\end{equation}
and $R$ is then defined by the position of the minimum of the total energy
per site in the first (Hartree) approximation \cite{LoHoFe06}.

\section{Theory}\label{sec:theory}

In this section we present our analytical results for the optical
conductivity. We first derive a general expression which allows us to
calculate the real part of the optical conductivity from the momentum and
energy dependent one-electron spectral function $A(k,\om)$. Approximations
for the spectral function in different coupling regimes have been obtained in
\cite{LoHoFe06}, and the basic formulas necessary for the computation of the
conductivity are compiled in \ref{sec:a}.  The calculations presented in
\cite{LoHoFe06} are based on the self-consistent treatment of the self-energy
equations in second-order perturbation theory. To describe the crossover from
the SC to the WC regime, we use the variational procedure outlined above.

Our approximation neglects vertex corrections due to direct fermion-fermion
interaction, which is also absent from the spinless
Hamiltonian~(\ref{eq:HM}).  Nevertheless, fermion-phonon and---at finite
carrier density---phonon-mediated, retarded fermion-fermion interaction
effects enter via the spectral functions \cite{LoHoFe06}. For simplicity, we
exclude from our discussion the quantum phase transition from a Luttinger
liquid to an insulating Peierls phase at half filling \cite{SyHuBeWeFe04,HoWeBiAlFe06}.

\subsection{General expression for the optical conductivity}

The current density operator for the 1D spinless Holstein model reads 
\begin{equation}\label{eq:currentop}
  \hat{\jmath}
  =
  \rmi
  \frac{e t}{V}
  \sum_{\las g',g\ras}
  (g'-g)
  c^\dag_{g'} c^{\nag}_{g}
  \,,
\end{equation}
where $g',g$ number the lattice sites, and $a=|g'-g|$ respectively $V=Na$ are
the lattice constant and the volume.

According to linear response theory \cite{Zubarev74}, the frequency dependent
complex conductivity $\sigma(\om)$ is determined by the Green function of the current
density operators. In particular, 
\begin{equation}\label{eq:resigma1}
  \Re\,\sigma(\om)
  =
  -\frac{V}{\om}\Im\,G^\mathrm{R}_{\hat{\jmath}}(\om)
  \,,
\end{equation}
where
\begin{equation}
  G^\mathrm{R}_{\hat{\jmath}}(\om)
  =
  \int_{-\infty}^\infty \rmd(t-t') G^\mathrm{R}_{\hat{\jmath}}(t-t')
  \rme^{\rmi\om(t-t')}
\end{equation}
with the retarded Green function
\begin{equation}
  G^\mathrm{R}_{\hat{\jmath}}(t-t')
  =
  -\rmi\las[{\hat{\jmath}}(t),{\hat{\jmath}}(t')]\ras \theta(t-t')
  \,.
\end{equation}
In the notation of \cite{Rick84}, the related Matsubara Green function 
takes the form
\begin{eqnarray}\label{eq:matsubara}\nonumber
  G_{\hat{\jmath}}(\tau_1,\tau_1')
  &=&
  -\las T_\tau {\hat{\jmath}}(\tau_1) {\hat{\jmath}}(\tau_1')\ras
  \\\nonumber
  &=&
  \left(\frac{et}{N}\right)^2\sum_{\las m_1,m_2\ras}\sum_{\las m_1',m_2'\ras}
  (m_1-m_2)(m_1'-m_2')
  \\\nonumber
  &&\times
  \las
  T_\tau c^\dag_{m_1}(\tau_1)c^{\nag}_{m_2}(\tau_1)c^\dag_{m_1'}(\tau_1')
  c^{\nag}_{m_2'}(\tau_1')
  \ras
  \\\nonumber
  &=&
  \left(\frac{et}{N}\right)^2
  \sum_{\las m_1,m_2\ras}\sum_{\las m_1',m_2'\ras}
  (m_1-m_2)(m_1'-m_2')
  \\
  &&\times\left.
  \las
  T_\tau c^{\nag}_{m_2}(\tau_2)c^{\nag}_{m_2'}(\tau_2')c^\dag_{m_1'}(\tau_1')
  c^\dag_{m_1}(\tau_1)
  \ras
  \right|_{{\tau_2=\tau_1-},\,{\tau_2'=\tau_1'-}}
  ,
\end{eqnarray}
where $m_i,m'_i$ are integers labeling the sites of the 1D lattice. The
two-particle Green function
\begin{equation}\label{eq:G4}
  G(2,2';1,1')
  =
  \las
  T_\tau c^{\nag}_{m_2}(\tau_2)c^{\nag}_{m_2'}(\tau_2')c^\dag_{m_1'}(\tau_1')
  c^\dag_{m_1}(\tau_1)
  \ras
\end{equation}
for mutually independent fermions can be written in terms of one-particle
Green functions \cite{Rick84}. Assuming $\las\hat{\jmath}\ras=0$ in
the absence of an electric field, the part of the two-particle Green function
relevant for the evaluation of equation~(\ref{eq:matsubara}) reads \cite{Rick84}
\begin{eqnarray}\label{eq:G2G2}\nonumber
  G(2,2';1,1')
  &=&
  -G(2,1')G(2',1)
  \\
  &=&
  -\las T_\tau c^{\nag}_{m_2}(\tau_2) c^\dag_{m_1'}(\tau_1')\ras
  \las T_\tau c^{\nag}_{m_2'}(\tau_2') c^\dag_{m_1}(\tau_1)\ras
  \,.
\end{eqnarray}
We use the Fourier transformation of the one-particle Green functions
\begin{equation}\label{eq:FT}
  G(2,1') 
  =
  \frac{1}{N}
  \sum_k
  \rme^{\rmi k(m_2-m_1')}
  \frac{1}{\beta}
  \sum_{\om_\nu}
  \rme^{-\rmi\om_\nu(\tau_2-\tau_1')}
  G(k,\rmi\om_\nu)
\end{equation}
with $\om_\nu=(2\nu+1)\pi/\beta$, $k=2\pi m/N$ (likewise for
$G(2',1)$), and the spectral representation
\begin{equation}\label{eq:spec}
  G(k,\rmi\om_\nu)
  =
  \int_{-\infty}^\infty
  \rmd \om'
  \frac{A(k,\om')}{\rmi\om_\nu-\om'}
  \,,
\end{equation}
to obtain
\begin{equation}\label{eq:gjint}
  G_{\hat{\jmath}}(\rmi\om_n)
  =
  \int_0^\beta \rmd (\tau_1-\tau_1') \rme^{\rmi\om_n(\tau_1-\tau_1')}
  G_{\hat{\jmath}}(\tau_1-\tau_1')
  \,,
\end{equation}
where $\om_n=2 n \pi/\beta$. Using equations~(\ref{eq:G4})--(\ref{eq:spec})
to express the rhs of equation~(\ref{eq:matsubara}), the Fourier
transformation~(\ref{eq:gjint}) is evaluated by carrying out the summations
over $m_i$, $m_i'$, and the integration and subsequent summation over the
Matsubara frequencies $\om_\nu$ \cite{Rick84}. We find
\begin{equation}
  \fl
  G_{\hat{\jmath}}(\rmi\om_n)
  =  
  \left(\frac{2et}{N}\right)^2
  \sum_k (\sin k)^2
  \int_{-\infty}^\infty \rmd \om' \int_{-\infty}^\infty \rmd \en'
  A(k,\om') A(k,\en')
  \frac{f(\om')-f(\en')}{\om'-\en'+\rmi \om_n}
  \,.
\end{equation}
Here $f(x)=[\exp(\beta x)+1]^{-1}$ is the Fermi function.
Using Dirac's identity 
\begin{equation}
  \frac{1}{\om+\om'-\en'+\rmi0^+}
  =
  \frac{\mathcal{P}}{\om+\om'-\en'}
  -
  \rmi\pi\delta(\om+\om'-\en')
  \,,
\end{equation}
the analytical continuation
$\rmi\om_n\mapsto\om+\rmi\delta$ gives in the limit $\delta\to0^+$
\begin{eqnarray}\nonumber
  \fl
  \Im\,G^\mathrm{R}_{\hat{\jmath}}(\om)
  =
  -\pi\left(\frac{2et}{N}\right)^2
  \sum_k (\sin k)^2
  \int_{-\infty}^\infty \rmd \om' \int_{-\infty}^\infty \rmd \en'
  A(k,\om') A(k,\en')\\
  \times
  [f(\om')-f(\en')]\delta(\om+\om'-\en')
  \,.
\end{eqnarray}
In particular, for $\om>0$ and $T\to0$, and introducing the Heaviside step
function $\theta(x)$,
\begin{equation}\label{eq:img}
  \fl
  \Im\,G^\mathrm{R}_{\hat{\jmath}}(\om)
  =
   -\pi\left(\frac{2et}{N}\right)^2
  \sum_k (\sin k)^2  \int_{-\infty}^0 \rmd \om' 
  A(k,\om') A(k,\om'+\om)
  \theta(\om'+\om)
  \,.
\end{equation}
The real part of $\sigma(\om)$ for $\om>0$, given by
\begin{equation}\label{eq:resigma}\fl
  \sigma^\mathrm{reg}(\om)
  =
  \frac{4\sigma_0}{\om}\frac{1}{N}\sum_k (\sin k)^2
  \int_{-\infty}^0\rmd\om'A(k,\om')A(k,\om+\om')\theta(\om+\om')
  \,,
\end{equation}
is determined by the overlap of the electronic spectral functions for energies
$\om'<0$ (relative to $\mu$) and $\om'+\om>0$, respectively. Here we have
defined $\sigma_0=a\pi(et)^2$.

The neglect of vertex corrections in the current-current response function
based on equation~(\ref{eq:G2G2}) leads to the factorisation into spectral
functions in equation~(\ref{eq:resigma}), which is equivalent to the exact
$\mathrm{D}=\infty$ result commonly used in dynamical mean-field theory
(DMFT) \cite{pruschke_review}. However, in contrast to the local
approximation of DMFT, the spectral functions used here have a
non-trivial momentum dependence \cite{LoHoFe06}.

\subsection{Limiting cases}

\paragraph{Weak coupling}

The explicit form of the WC electronic spectral function,
calculated according to equations~(\ref{eq:A1})--(\ref{eq:zk}), was
determined in \cite{LoHoFe06}. Taking into account the condition $|\om|<\om_0$
($|\om|>\om_0$) for the coherent (incoherent) part, and the explicit
result~(\ref{eq:sc_coherent}) for the coherent part $A^\mathrm{c}(k,\om)$, we
find for the optical conductivity
\begin{eqnarray}\fl\nonumber\label{eq:oc_wc}
  \sigma^\mathrm{reg}(\om)
  =
  \frac{4\sigma_0}{\pi\om}\int_0^\pi \rmd k (\sin k)^2
  \\\nonumber\hspace*{-2em}
  \times\Big[
   z_k\theta(-(\Ek-\mu))\theta(\om_0+\Ek-\mu)\theta(\Ek-\mu+\om-\om_0)
  A^\mathrm{ic}(k,\Ek-\mu+\om)
  \\\nonumber\hspace*{-1em}
  +z_k\theta(\Ek-\mu)\theta(\om_0-(\Ek-\mu))\theta(\om-(\Ek-\mu)-\om_0)
  A^\mathrm{ic}(k,\Ek-\mu-\om)
  \\\hspace*{-1em}
  +\int_{-\infty}^0\rmd\om' A^\mathrm{ic}(k,\om')A^\mathrm{ic}(k,\om+\om')
  \theta(-\om'-\om_0)\theta(\om+\om'-\om_0)\Big]
  \,.
\end{eqnarray}

\begin{figure}
  \centering
  \includegraphics[width=0.4\textheight]{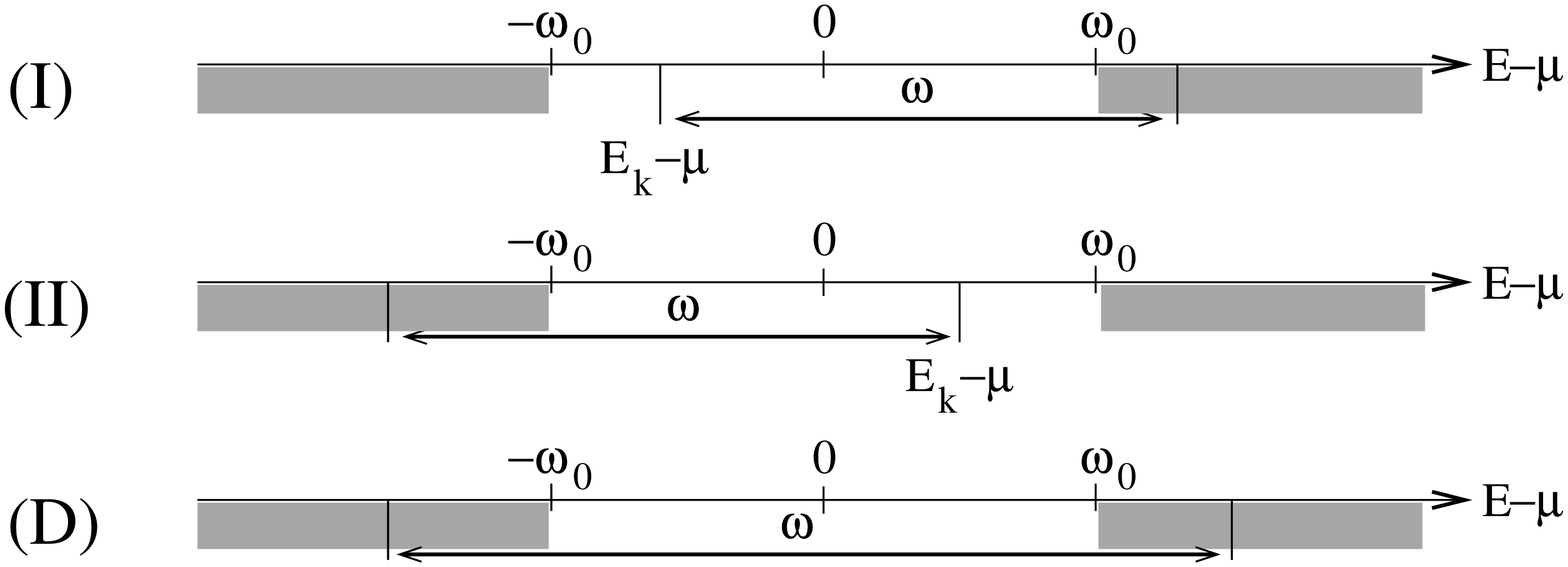}
  \caption{\label{fig:trans_wc}    
  Illustration of the transitions contributing to
  equation~(\ref{eq:oc_wc}).}
\end{figure}

For the discussion of equation~(\ref{eq:oc_wc}) in section~\ref{sec:results},
we have to recall \cite{LoHoFe06} that the WC approximation of the incoherent one-electron
spectral function $A^\mathrm{ic}(k,\om')$ is non-zero only for
$\om'\in(-\om_0-W-\mu,-\om_0)$ or $\om'\in(\om_0,\om_0+W-\mu)$. The last
term in equation~(\ref{eq:oc_wc}), corresponding to transitions between these
frequency intervals of the spectrum, is denoted in figure~\ref{fig:trans_wc}
as $D$.

\paragraph{Intermediate coupling}

Within the variational approach, the electronic spectral function is given by
equation~(\ref{eq:sc_el_spectrum}) with $g$ replaced by $\gammab g$ and
$\tilde{W}=W\rme^{-\gammab^2 g^2}$. The explicit expression for
$\Sigma(k,\om)$ determining the polaronic spectral function can be found in
\cite{LoHoFe06}. Using the definitions~(\ref{eq:Ae<})--(\ref{eq:abcddef}), we
can write
\begin{equation}\label{eq:ABCD}
  \int_{-\infty}^0 \rmd \om' \theta(\om+\om')
  \Ae(k,\om') \Ae(k,\om'+\om) = A + B + C + D
  \,.
\end{equation}
In general all terms of equation~(\ref{eq:ABCD}) contribute to
$\sigma^\mathrm{reg}(\om)$. We find
\begin{eqnarray}\fl\nonumber\label{eq:oc_sc}
  [\sigma^\mathrm{reg}(\om)]_A
  =  
  \frac{4\sigma_0}{\pi^2\om} \rme^{-2(\gammab g)^2}
  \sum_{s\geq1}\frac{(\gammab g)^{2s}}{s!}\int_0^\pi\rmd k (\sin k)^2 z_k
  \\\nonumber
  \times\Big[
  \frac{z_{k_1}}{|\partial_{k_1} E_{k_1}|}
  \theta(\Ek+\eta)\theta(\om_0-(\Ek+\eta))\theta(-E_{k_1}-\eta)\theta(\om_0+E_{k_1}+\eta)
  \\\nonumber
  +
  \frac{z_{k_2}}{|\partial_{k_2}E_{k_2}|}
  \theta(-\Ek-\eta)\theta(\om_0+\Ek+\eta)\theta(E_{k_2}+\eta)\theta(\om_0-(E_{k_2}+\eta))
  \Big]\\\nonumber
  \hspace*{-2em}+
  \frac{2\sigma_0}{\om\pi^2} \rme^{-2(\gammab g)^2}
  \sum_{s\geq1}\frac{2^s(\gammab g)^{2s}}{s!}
  (1-2^{1-s})
  \int_0^\pi\rmd k z_k \frac{z_{k'}}{|\partial_{k'}E_{k'}|}
  \\
  \times  
  \theta(-\Ek-\eta)\theta(\om_0+\Ek+\eta)
  \theta(E_{k'}+\eta)\theta(\om_0-(E_{k'}+\eta))
  \,,
\end{eqnarray}
where
\begin{eqnarray}\label{eq:zkEk}\fl\nonumber
 \left[\frac{z_{k}}{|\partial_k E_k|}\right]^{-1}
  =
  \Bigg|
  \tilde{W}\sin k
  \Bigg\{
    1 - \frac{2}{\pi}\cos k \sum_{s\geq1}\frac{(\gammab g)^{2s}}{s!}\mathcal{P}
    \int_{-\tilde{W}}^{\tilde{W}}\frac{\rmd\xi}{\sqrt{1-(\xi/\tilde{W})^2}}
  \\\nonumber
  \times  
  \left[
    \frac{\theta(\xi+\eta)}{E_k-s\om_0-\xi}
    +
    \frac{\theta(-\xi-\eta)}{E_k+s\om_0-\xi}
  \right]
  \Biggr\}
  +
  \frac{2}{\pi}\gammab(1-\gammab)\Ep\sin k
  \\
  \times
  \mathcal{P}
  \int_{-\tilde{W}}^{\tilde{W}}\frac{\rmd\xi}{\sqrt{1-(\xi/\tilde{W})^2}}
  \left[
    \frac{\theta(\xi+\eta)}{E_k-\om_0-\xi}
    -
    \frac{\theta(-\xi-\eta)}{E_k+\om_0-\xi}
  \right]
  \Bigg|
  \,.
\end{eqnarray}
Here $k_1,k_2\in(0,\pi)$ are determined by the conditions
$E_{k_1}=\Ek-(\om-s\om_0)$ and
$E_{k_2}=\Ek+(\om-s\om_0)$, whereas $k'>k_\mathrm{F}$ is given as the solution of
$E_{k'}=\Ek+(\om-s\om_0)$. 

For the other contributions to equation~(\ref{eq:ABCD}), we have
\begin{eqnarray}\fl\nonumber\label{eq:B}
 B 
 =
 \rme^{-(\gammab g)^2}z_k \Ae^\mathrm{ic}(k,\om+\Ek+\eta)\theta(\om+\Ek+\eta)
 \theta(-\Ek-\eta)\theta(\om_0+\Ek+\eta)
 \\\nonumber
 +\rme^{-(\gammab g)^2}\sum_{s\geq1}\frac{(\gammab g)^{2s}}{s!}\frac{1}{\pi}
 \int_0^\pi\rmd k' z_{k'}
 \Ae^\mathrm{ic}(k,\om+E_{k'}+\eta-s\om_0)
 \\
 \times\theta(\om+E_{k'}+\eta-s\om_0)
 \theta(-E_{k'}-\eta)\theta(E_{k'}+\eta+\om_0)
 \,,
\end{eqnarray}
\begin{eqnarray}\fl\nonumber\label{eq:C}
 C 
 =
 \rme^{-(\gammab g)^2}z_k \Ae^\mathrm{ic}(k,\Ek+\eta-\om)\theta(\om-\Ek-\eta)
 \theta(\Ek+\eta)\theta(\om_0-\Ek-\eta)
 \\\nonumber
 +\rme^{-(\gammab g)^2}\sum_{s\geq1}\frac{(\gammab g)^{2s}}{s!}\frac{1}{\pi}
 \int_0^\pi\rmd k' z_{k'}
 \Ae^\mathrm{ic}(k,E_{k'}+\eta+s\om_0-\om)
 \\
 \times
 \theta(\om-E_{k'}-\eta-s\om_0)
 \theta(E_{k'}+\eta)\theta(\om_0-E_{k'}-\eta)
 \,,
\end{eqnarray}
and
\begin{equation}\label{eq:D}
 D 
 =
 \int_{-\infty}^0\rmd\om' \Ae^\mathrm{ic}(k,\om')\Ae^\mathrm{ic}(k,\om+\om')
 \theta(\om+\om')
 \,,
\end{equation}
so that
\begin{equation}
 [\sigma^\mathrm{reg}(\om)]_{B+C+D}
  =  
  \frac{4\sigma_0}{\pi\om}\int_0^\pi \rmd k (\sin k)^2 (B+C+D)
  \,.
\end{equation}
The parts of the electronic spectral function denoted by
$A^\mathrm{ic}_\mathrm{e}(k,\om')$, defined in equation~(\ref{eq:app:aeic}),
are determined by the incoherent parts of the polaronic spectral function.

\paragraph{Strong coupling}

For $g^2\gg1$, small polarons are the correct fermionic quasiparticles, and
the calculations corresponding to equations~(\ref{eq:A1})--(\ref{eq:zk}) are
done using the small-polaron Hamiltonian defined by the
coefficients~(\ref{eq:scC}) (see \cite{LoHoFe06} for explicit results).  In
this regime, the zeroth-order small-polaron approximation---with the spectrum
consisting only of coherent states in the polaron band---is applicable and
the polaronic spectral weights $z_k\approx1$ \cite{LoHoFe06}. Consequently,
it is sufficient to keep only the term $A$, and SC expressions corresponding to
equations~(\ref{eq:oc_sc}) and~(\ref{eq:zkEk}) can be obtained by setting
$\gammab=1$.

\begin{figure}
  \centering
  \includegraphics[width=0.4\textheight]{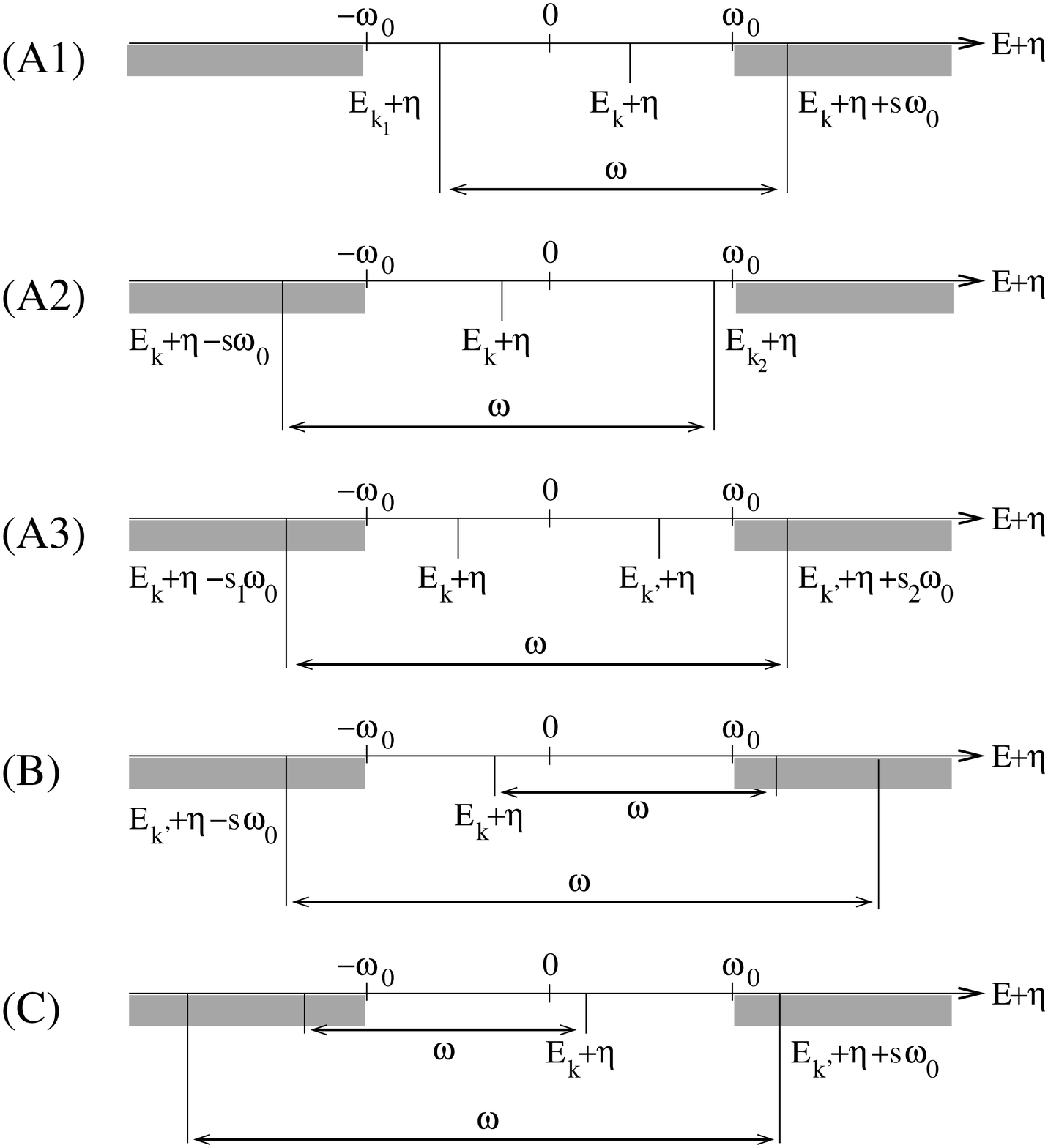}
  \caption{\label{fig:trans_ic}    
  Transitions contributing to equations~(\ref{eq:oc_sc}), (\ref{eq:B})
  and~(\ref{eq:C}). Equation~(\ref{eq:D}) corresponds to
  figure~\ref{fig:trans_wc}(D) with $-\mu$ replaced by $\eta$.}
\end{figure}

The rather complicated explicit formulas for the optical conductivity
presented above are
visualised in figures~\ref{fig:trans_wc} and~\ref{fig:trans_ic} for the WC
and the SC/IC cases, respectively. In particular, these pictures show that
non-zero values of $\sigma^\mathrm{reg}(\om)$ are restricted to
$\om>\om_0$. The existence of such a threshold for the optical conductivity
at $T=0$ was already recognised in \cite{GuLaFi62} on the basis of WC
calculations in the framework of the continuous polaron model.

\section{Results}\label{sec:results}

We now come to a discussion of numerical results obtained from the
analytical expressions derived in section~\ref{sec:theory}, which will be
compared to exact data for one electron. It is instructive to first consider
the kinetic energy and the Drude weight, related to
$\sigma^\mathrm{reg}(\om)$ by the f-sum rule.

\begin{figure}
  \centering
  \includegraphics[height=0.35\textwidth]{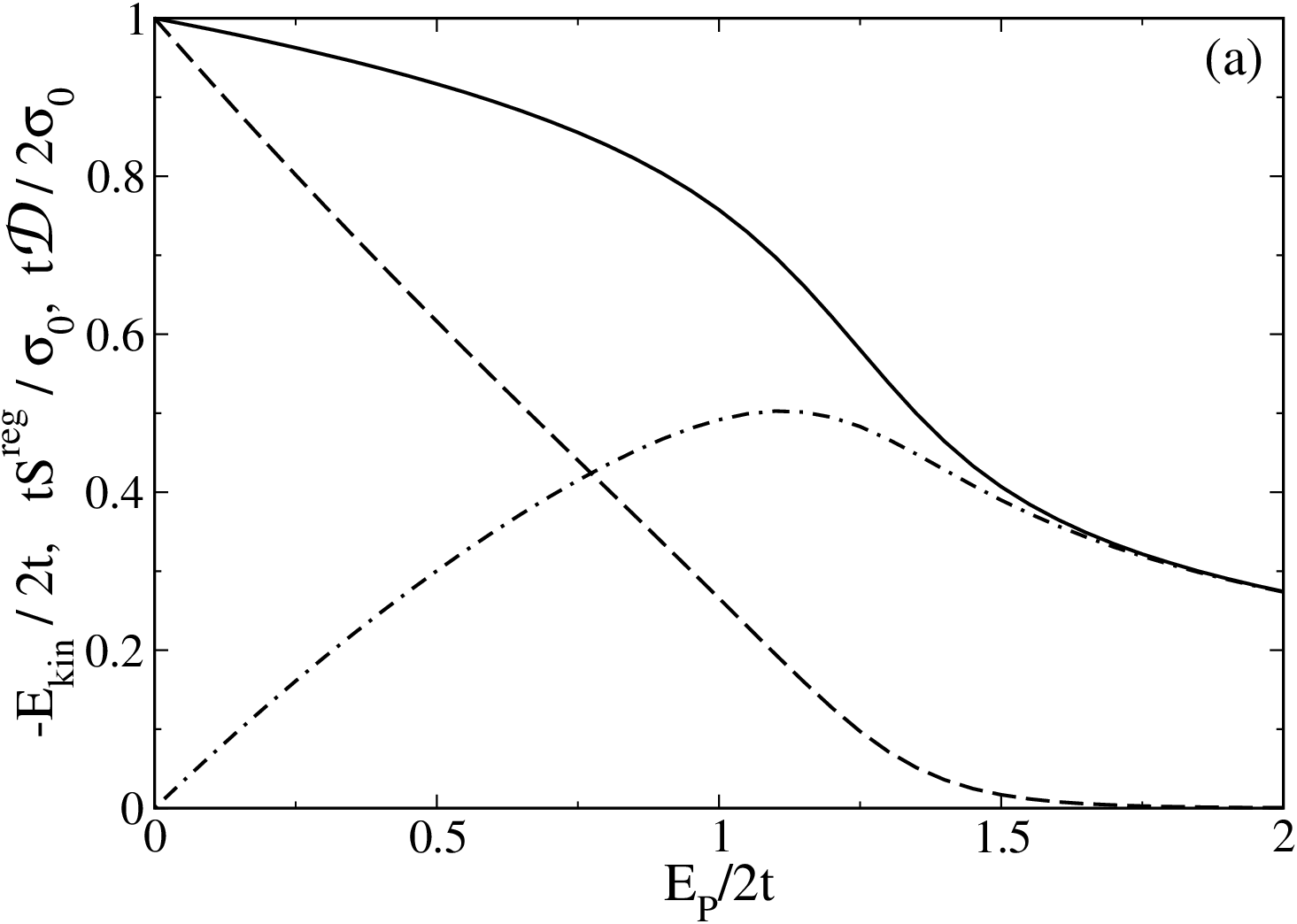}
  \includegraphics[height=0.35\textwidth]{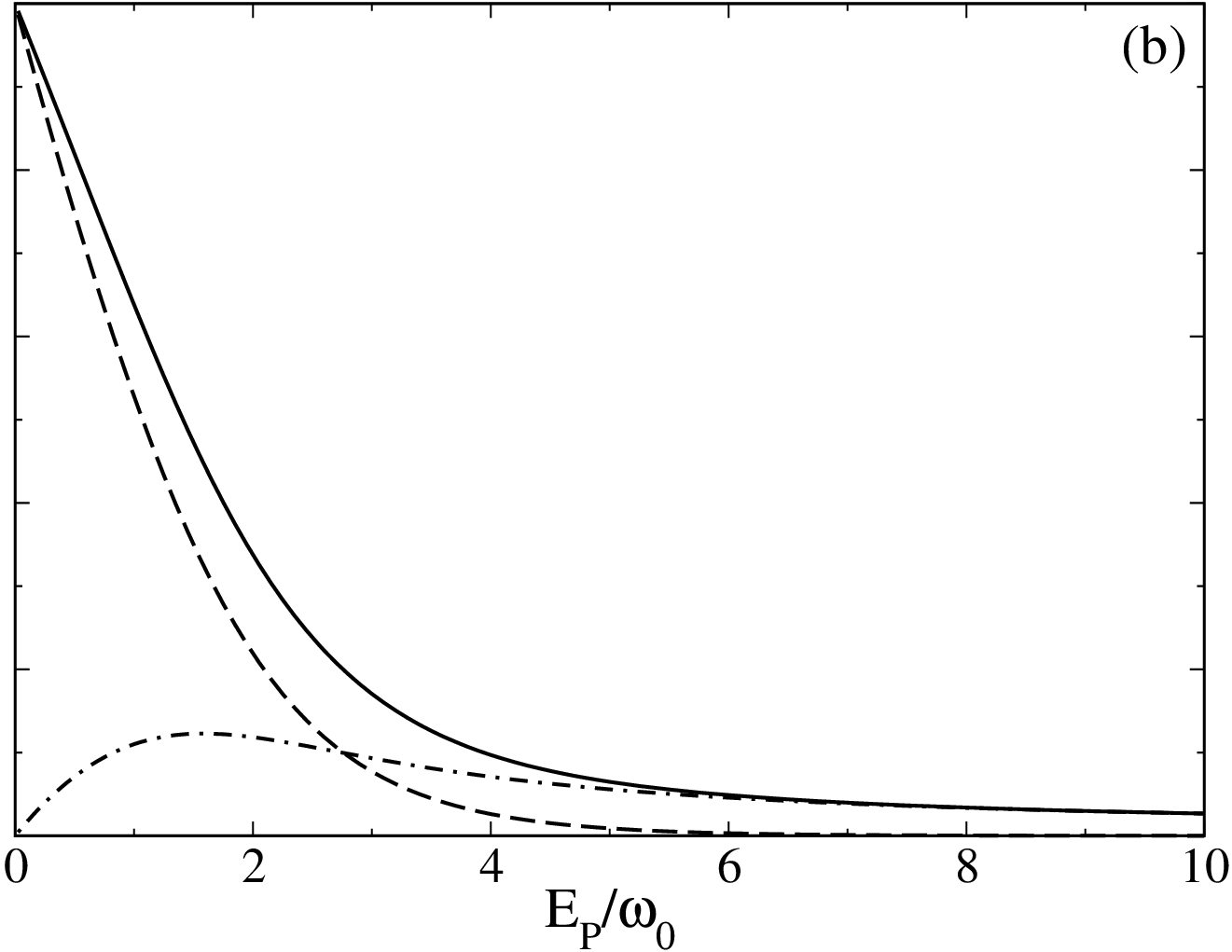}
  \caption{\label{fig:vd_drudesumrule} Numerical results for the kinetic
    energy $E_\mathrm{kin}$ ($\full$), the Drude weight $\mathcal{D}$
    ($\dashed$), and the integrated spectral weight $\Sr$ ($\chain$,
    equation~(\ref{eq:S_reg})), for one electron as a function of
    electron-phonon coupling strength. Here (a) $\om_0/t=0.4$ and (b)
    $\om_0/t=4$.}
\end{figure}

\subsection{Integral quantities}

In a tight-binding system with the kinetic energy operator
\begin{equation}
  \hat{T} = -t\sum_{\las i,j\ras} c^\dag_i c^{\nag}_j
\end{equation}
and the current density operator~(\ref{eq:currentop}), it follows that
\cite{Ma77,BaGrRi87}
\begin{equation}\label{eq:sumrulegen}
  -e^2 a\varepsilon_\mathrm{kin}
  =
  \frac{1}{\pi}\int_{-\infty}^\infty \rmd \om \sigma(\om)
  =
  \frac{2}{\pi}\int_{0}^\infty \rmd \om \Re\,\sigma(\om)
  \,,
\end{equation}
by symmetry of $\sigma(\om)$. Here,
$\varepsilon_\mathrm{kin}=\las \hat{T}\ras/N$ represents the kinetic energy per
lattice site. 

For $\om>0$ and $T=0$, the electron-phonon interaction implied in the
Holstein model considered here gives a non-zero $\Re\,\sigma(\om)$ only if
$\om>\om_0$. At $\omega=0$, $\Re\,\sigma(\om)$ has a singularity to be
deduced in \ref{sec:b}. Writing
\begin{equation}\label{eq:re_sigma}   
  \Re\,\sigma(\omega)
  =
  \mathcal{D} \delta(\omega)+ \sir(\omega)
  \,,
\end{equation}
where $\mathcal{D}$ is the so-called Drude weight, and defining
$\Sr\equiv\Sr(\infty)$ with
\begin{equation}\label{eq:S_reg}
  \Sr(\omega)
  =
  \int_{0}^\omega\rmd \omega' \sir(\omega')
  \,,
\end{equation} 
the f-sum rule~(\ref{eq:sumrulegen}) takes the form
\begin{equation}\label{eq:sumrulegen2}
  -e^2 a\varepsilon_\mathrm{kin}
  =
  \frac{1}{\pi}\mathcal{D}
  +
  \frac{2}{\pi} S^\mathrm{reg}
  \,.
\end{equation}

The kinetic energy may be determined independent of the rhs of
equation~(\ref{eq:sumrulegen2}) from the electronic spectral function as
\cite{SePePl02}
\begin{equation}\label{eq:ekin_Akw}
  \varepsilon_\mathrm{kin}
  =
  \frac{1}{N}
  \sum_k
  \int_{-\infty}^0\rmd\om\,  
  \varepsilon_k\,A(k,\om)
  \,,\quad
  \varepsilon_k = -2t \cos k
  \,.
\end{equation}

In the case of one electron in a lattice with $N$ sites (\ie, with carrier
concentration $n=1/N$), $\varepsilon_\mathrm{kin}$ is equal to the kinetic
energy $E_\mathrm{kin}$ of one electron divided by $N$. To obtain non-zero
results, both sides of equation~(\ref{eq:sumrulegen2})
are multiplied by $N$ before taking the limit $N\to\infty$. Accordingly, using
the spectral representation at $T=0$, we have for $\om>0$
\begin{equation}\label{eq:sigma_reg}
  \sir(\omega) 
  = 
  \frac{a\pi}{\omega} \sum_{E_m>E_0} 
  |\langle \psi_m|\hat{\jmath}|\psi_0\rangle |^2\ 
  \delta[\omega - (E_m - E_0)]
\end{equation}
with the current operator $\hat{\jmath} = - \rmi e t\sum_{i}(c_{i}^{\dagger}
c_{i+1}^{} - c_{i+1}^{\dagger} c_{i}^{})$ for a single electron. Here
$|\psi_m\rangle$ denotes the $m$-th excited state with energy $E_m$.
In comparison to equations~(\ref{eq:currentop}),~(\ref{eq:resigma1}) a factor
$1/N$ is omitted in equation~(\ref{eq:sigma_reg}). 
   
According to Kohn's formula, the Drude weight of the Holstein model with one
electron can be calculated from the effective mass $m^*$ as
($E_k$ is the polaron band dispersion) \cite{PhysRev.133.A171}
\begin{equation}\label{eq:drudemass}
  \mathcal{D}
  =e^2\pi a
  \left.\partial_k^2 E_k\right|_{k=0}
  =e^2\pi a\frac{1}{m^*}
  \,.
\end{equation} 
The evaluation of $\mathcal{D}$ via the f-sum rule, and the determination of
$m^*$ are independent computations, so that equation~(\ref{eq:drudemass}) may
be used to validate the numerics.

\begin{table}
  \centering
  \caption{\label{tab:Ekin}
    Analytical results for the kinetic energy $-\varepsilon_\mathrm{kin}/t$ for $\om_0/t=0.4$, QMC
    data for $N=16$, $\beta t=10$, $\Delta\tau=0.05$
    \cite{HoNevdLWeLoFe04}, and ED data for $N=10$ ($^*$--not fully converged
    with respect to the number of phonons $M=25$) \cite{WeBiHoScFe05}.}
  \vspace*{1em}
  \begin{tabular}{c|c|c|c}\hline\hline
    Parameters                         & Analytical    & QMC       & ED \\
    \hline
     $\Ep/t=0.1$, $n=0.1$  & 0.202 (WC)  0.211 (IC) & 0.190 & 0.1986\\
     $\Ep/t=0.1$, $n=0.3$  & 0.509 (WC)  0.518 (IC) & 0.504 & 0.5208\\
     $\Ep/t=0.1$, $n=0.4$  & 0.600 (WC), 0.609 (IC) & 0.596 & 0.5824\\
    \hline
     $\Ep/t=2.0$, $n=0.1$  & 0.155 (IC) & 0.136$\pm$0.002   &  0.1517\\
     $\Ep/t=2.0$, $n=0.3$  & 0.382 (IC) & 0.36 $\pm$0.01    &  0.4070\\
     $\Ep/t=2.0$, $n=0.4$  & 0.397 (IC) & 0.43 $\pm$0.02    &  0.4531\\
    \hline
     $\Ep/t=4.0$, $n=0.4$  & 0.13  (SC) & 0.19$\pm$0.01     &  \hspace*{0.5em}0.2318$^*$\\    
    \hline    
  \end{tabular}\\
\end{table}

\subsubsection{One electron}

We adopt the basis construction of \cite{BoTrBa99}, which allows for the
calculation of results that are variational in the thermodynamic limit, and
find that the f-sum rule is fulfilled to at least six digits for the
parameters shown. For spectral properties a Chebyshev expansion method is
used \cite{WeWeAlScFe05}.

The dependence of $E_\mathrm{kin}$, $\Sr$ and $\D$ on the electron-phonon
coupling strength reflects the well-known crossover from a large polaron at
weak coupling to a small polaron at strong coupling \cite{FeTr2006}. As
expected, the results in figure~\ref{fig:vd_drudesumrule} reveal a
significant dependence on the adiabaticity ratio $\om_0/t$. The adiabatic regime
(figure~\ref{fig:vd_drudesumrule}(a)) is characterised by rather pronounced
decrease of $E_\mathrm{kin}$ and $\D$ (increase of $m^*$) in the vicinity of
the point $\lambda=1$ near which the crossover occurs, which is compensated
by an increase of $\Sr$ due to enhanced incoherent scattering. In contrast,
in the non-adiabatic regime (figure~\ref{fig:vd_drudesumrule}(b)), these
changes occur over a much larger range of the relevant coupling constant
$g^2$. 

\subsubsection{Many electrons}

Equation~(\ref{eq:ekin_Akw}) turns out to be numerically problematic in the
SC regime, and $\varepsilon_\mathrm{kin}$ is instead calculated from
equation~(\ref{eq:app:ekin_sreg}). An approximation for $\mathcal{D}$ is
given by equation~(\ref{eq:app:drude_gen}).

To test the reliability of the spectral functions obtained from the
analytical approach we compare in table~\ref{tab:Ekin} the analytical kinetic
energy $\varepsilon_\mathrm{kin}$ for different parameter sets to numerical
data from quantum Monte Carlo (QMC) simulatons \cite{HoNevdLWeLoFe04} and exact
diagonalisation (ED) \cite{WeBiHoScFe05}. We restrict ourselves to the
adiabatic regime since the analytical approach generally works better for
$\om_0/t\gg1$ \cite{LoHoFe06}. The agreement is satisfactory in all cases,
with the deviations of the analytical results for $\Ep/t=2$ originating to
some degree in the missing spectral weight in $A(k,\om)$ (see below).  The SC
approach slightly underestimates the kinetic energy for $\Ep/t=4$, a coupling
which does not fall into the true SC regime (see also
figure~\ref{fig:vd_adiab}(c)).

\begin{figure}
  \centering
  \includegraphics[height=0.35\textwidth]{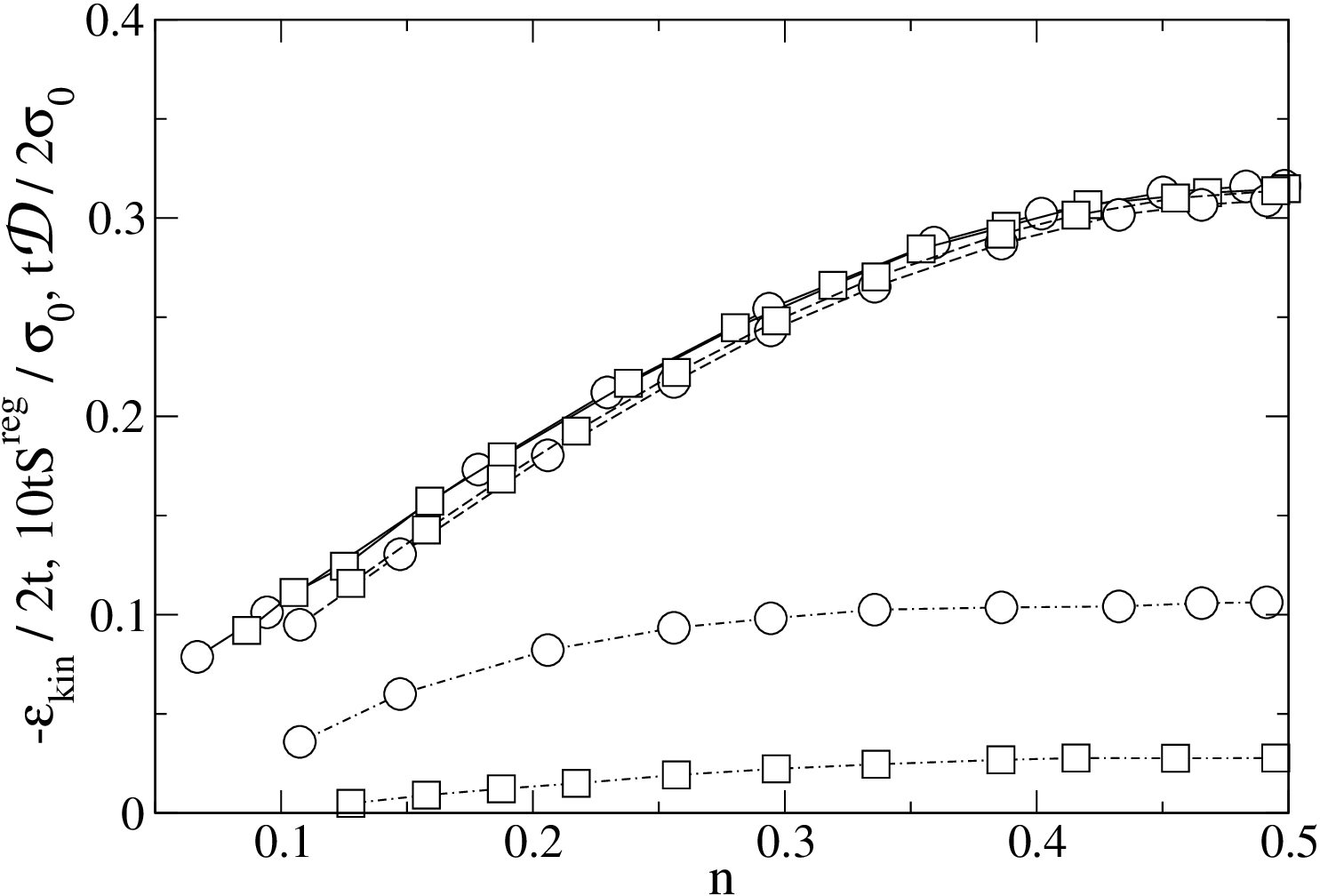}
  \caption{\label{fig:an_sumrule}
  Analytical results for the kinetic energy $\varepsilon_\mathrm{kin}$
  ($\full$), the Drude weight $\mathcal{D}$ ($\dashed$),  and the integrated
  spectral weight $\Sr$ ($\chain$) as a function of band filling $n$. Here $\Ep/t=0.1$ and
  $\om_0/t=0.4$ ($\opencircle$) respectively $\om_0/t=4$ ($\opensquare$).}
\end{figure}

\begin{figure}
  \centering
  \includegraphics[width=0.55\textwidth]{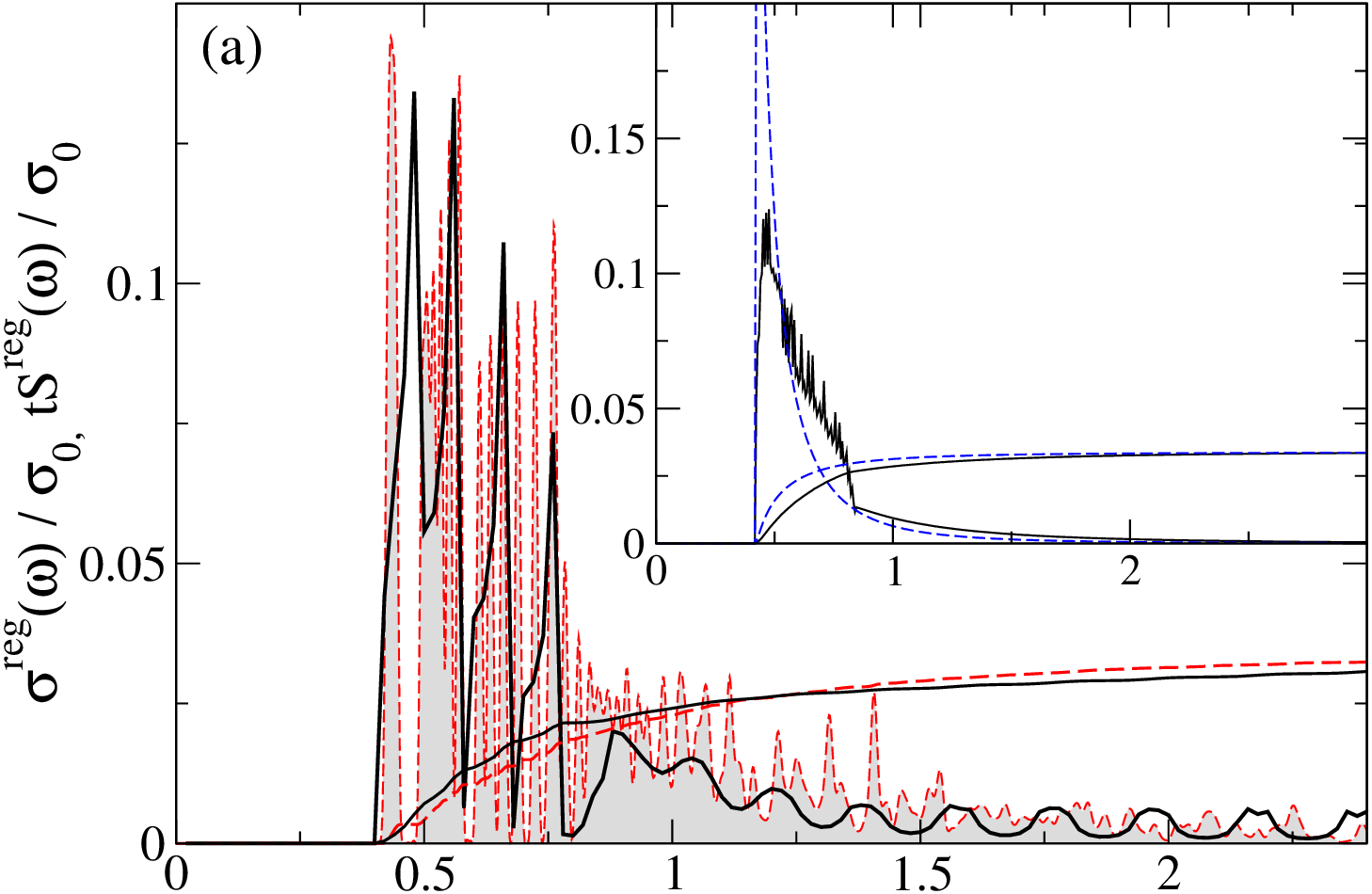}\\
  \includegraphics[width=0.55\textwidth]{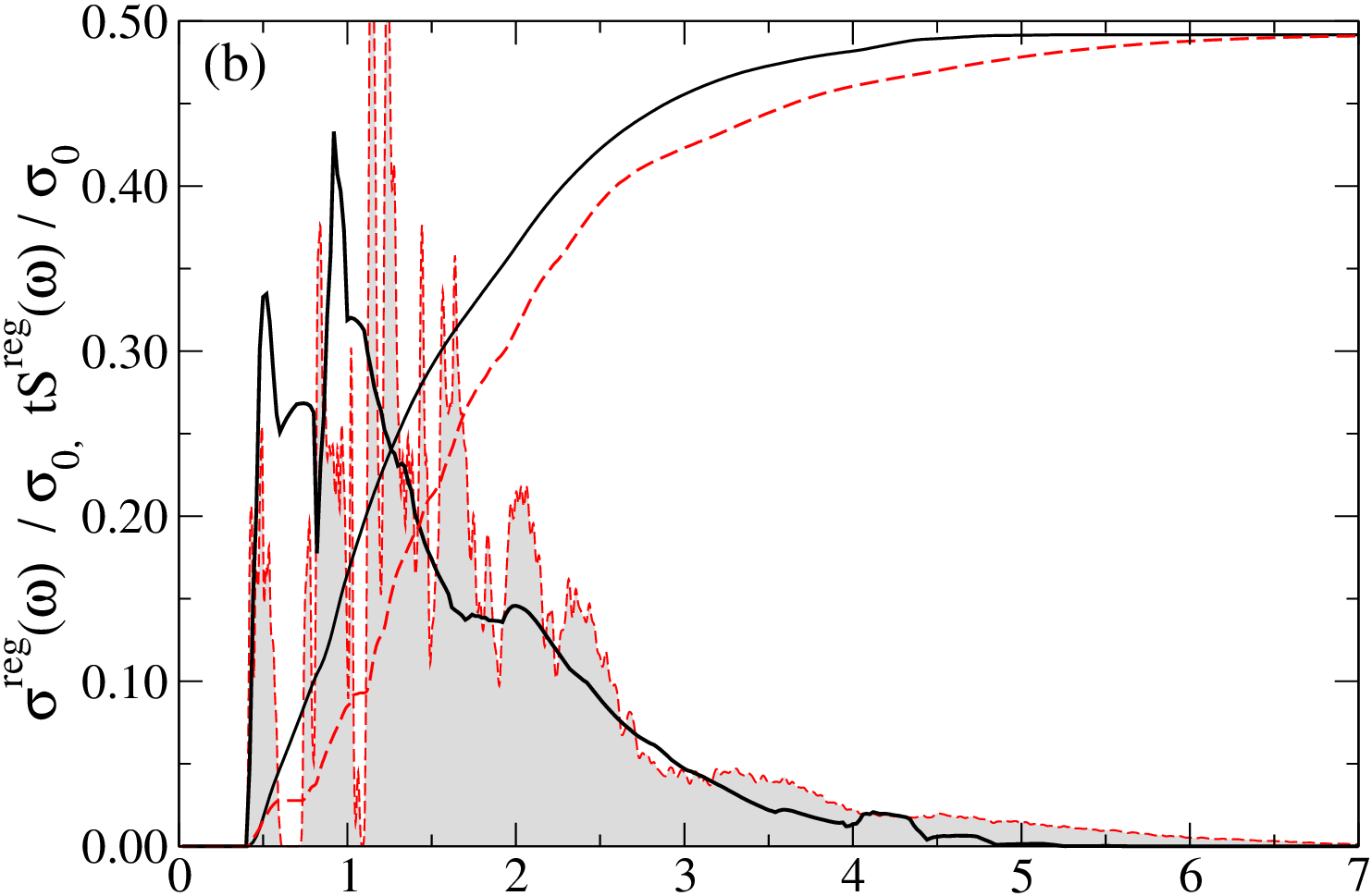}\\
  \includegraphics[width=0.55\textwidth]{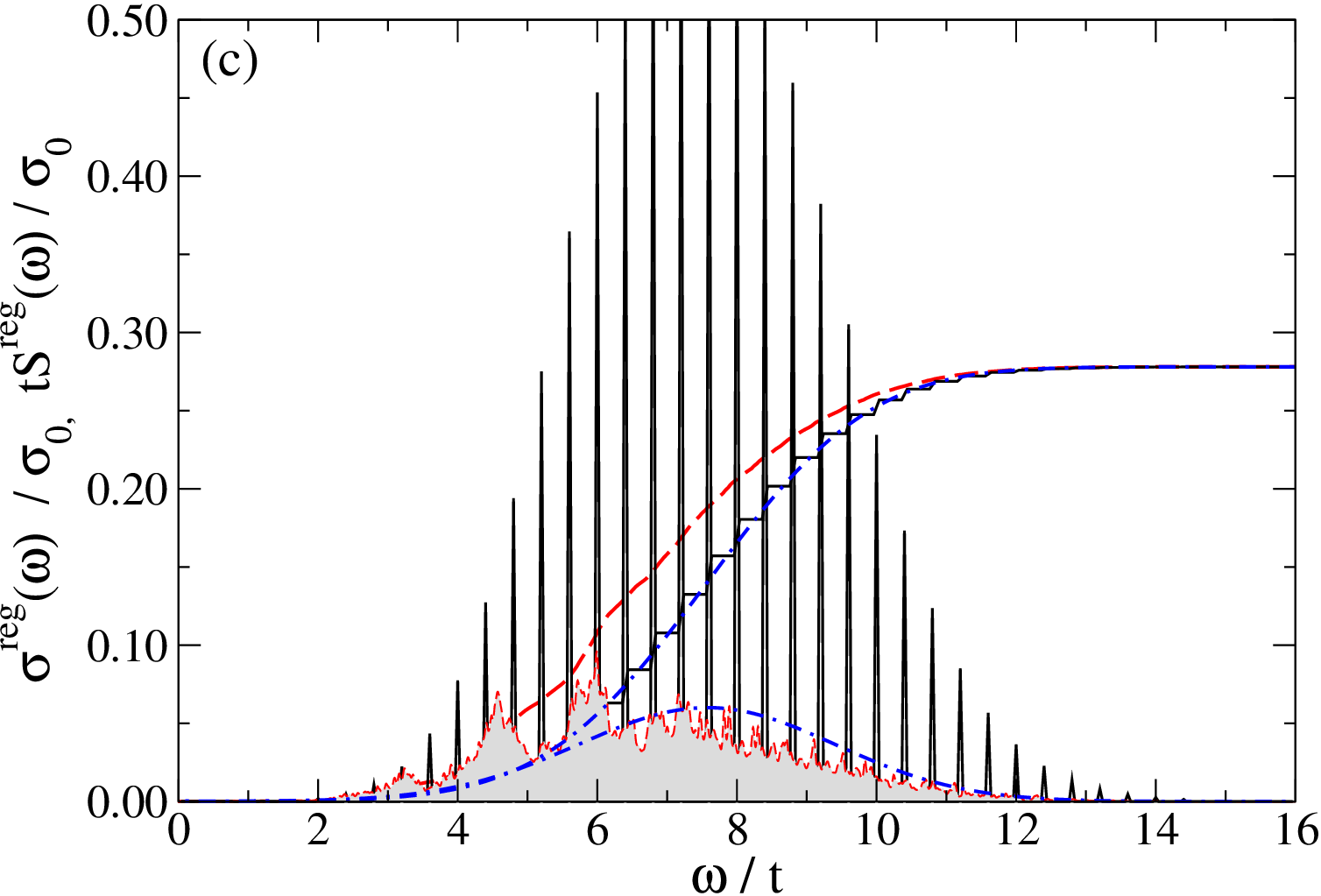}
  \caption{\label{fig:vd_adiab} (colour online) Numerical one-electron
    results ($\dashed$, red) for the regular part of the optical conductivity
    $\sir(\om)$ and the integrated spectral weight $\Sr(\om)$, and analytical
    results ($\full$, black, scaled to give the same $\Sr$) for $n=0.1$ from
    the IC approximation with (a) $R=2.4$, (b) $R=2.1$, (c) $R=0.1$.  Here
    $\om_0/t=0.4$ and (a) $\Ep/t=0.1$, (b) $\Ep/t=2$, and (c) $\Ep/t=4$. The
    inset in (a) compares the WC analytical result for $n=0.1$ ($\full$,
    black) to that obtained for $n\to0$ ($\dashed$, blue, see text), both
    rescaled to $S^\mathrm{reg}$ for one electron. Panel~(c) includes the
    small-polaron result of \cite{Em93} ($\chain$, blue).}
\end{figure}

Figure~\ref{fig:an_sumrule} shows the dependence of
$\varepsilon_\mathrm{kin}$, $\mathcal{D}$, and $\Sr$ on carrier density for
$\Ep/t=0.1$. Whereas the kinetic energy and the Drude weight are almost
identical in the adiabatic and non-adiabatic regime, the regular part $\Sr$
is substantially larger for $\om_0/t<1$ owing to enhanced incoherent
scattering of carriers by phonons, as is also observed in the one-electron
case (figure~\ref{fig:vd_drudesumrule}(a)). The weak coupling leads to very
small $\Sr$ (note the different scale used for $\Sr$) and hence
requires---via the f-sum rule---that $-\varepsilon_\mathrm{kin}/t\approx
t\mathcal{D}/\sigma_0$. A comparison of $\Sr$ to exact IC results
\cite{WeBiHoScFe05} reveals similar deviations as for
$\varepsilon_\mathrm{kin}$.

The approximations for the Drude weight $\mathcal{D}$ derived in~\ref{sec:b}
yield reliable results in the WC and SC cases, whereas the dependence of
$\mathcal{D}$ on $\Ep$ for intermediate coupling is not properly described.
Alternatively, an estimate for $\mathcal{D}$ may be obtained from the sum
rule, but will be affected by the deviations of $\varepsilon_\mathrm{kin}$
and $S^\mathrm{reg}$ from exact results.

\subsection{Optical response}\label{sec:results:opt}

\begin{figure}
  \centering
  \includegraphics[height=0.35\textwidth]{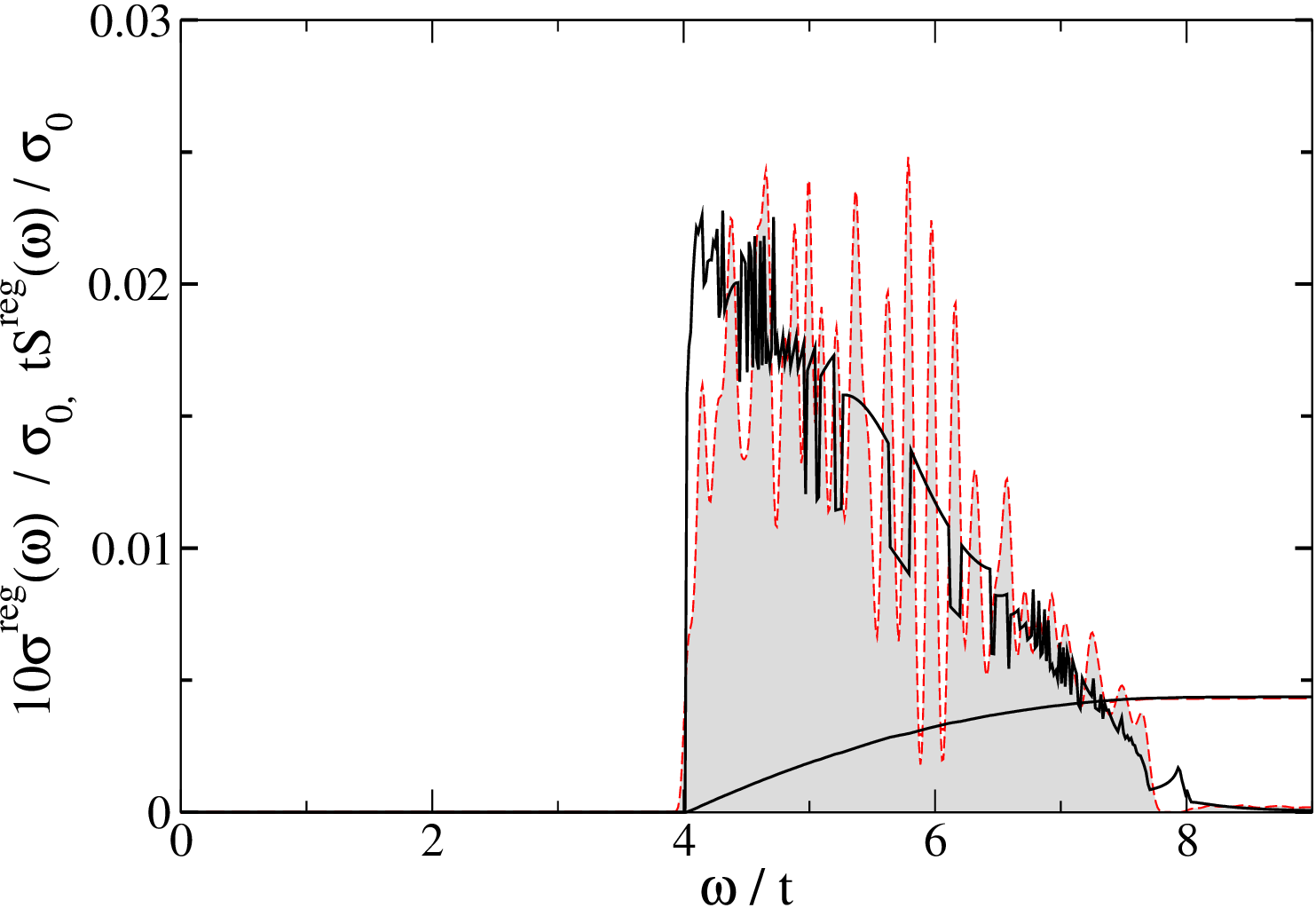}
  \caption{\label{fig:vd_nonadiab} (colour online) Numerical results for
    $\sir(\om)$ for one electron ($\dashed$, red), and analytical results
    ($\full$, black) for $n=0.1$ (scaled to the same $\Sr$) from the WC
    approximation. Here $\om_0/t=4$ and $\Ep/t=0.1$.  Also shown is
    $\Sr(\om)$.}
\end{figure}

\subsubsection{One electron}

Although the optical conductivity for one electron has been discussed
before \cite{FeTr2006}, we shall begin with this case as it provides a
framework for the finite-density case, and another test for
the analytical theory. We focus on the adiabatic regime $\om_0/t<1$ relevant
for many polaronic materials.

Figure~\ref{fig:vd_adiab} shows numerical and analytical results for
$\sir(\om)$ for $\om_0/t=0.4$ and three different values of $\Ep$. In order
to compare with the one-electron case, we have chosen $n=0.1$ in the
analytical approach, and scaled the results to yield the exact one-electron
value for $\Sr$. The dependence of $\sir(\om)$ and $\Sr$ on $n$
will be discussed below.

As anticipated from the analytical results (cf equations~(\ref{eq:oc_wc})
and~(\ref{eq:oc_sc})), there exists a minimal absorption energy
(absorption threshold) $\om=\om_0$. Another striking feature is the change from an
asymmetric spectrum in the large polaron regime to a (more) symmetric
spectrum in the small-polaron regime also observed experimentally
\cite{HaMaLoKo04}.

Starting with weak coupling $\Ep=0.1$ in figure~\ref{fig:vd_adiab}(a), we
find a pronounced absorption signal near $\om=\om_0$, followed by a
continuous decrease. The analytical results from the IC approximation
reproduce the main features (notice the good agreement for $\Sr(\om)$), with
the sharp low-energy peaks smeared out.  As illustrated by the WC results in
the inset (in good agreement with the IC results), this smearing depends on
the number of $k$-values used in the integration ($N=31$ and 801,
respectively). The inset also contains analytical results for $\mu\to E_0^+$,
corresponding to the one-electron case \cite{LoHoAlFe06}, for which a
non-zero WC result is obtained by using the zero-density spectral function in
equation~(\ref{eq:oc_wc}). Obviously, small peaks in
$\sigma^\mathrm{reg}(\om)$ are further washed out in this limit. The sharp
peaks in the exact numerical results are due to the use of a finite number of
poles in equation~(\ref{eq:sigma_reg}).

Turning to the IC results ($\Ep/t=2$, close to the critical coupling for the
small-polaron crossover) depicted in figure~\ref{fig:vd_adiab}(b), we find
low-energy peaks split off from the high-energy part and separated by
$\om_0$. Besides, the long high-energy tail of both the exact and analytical
curves is reminiscent of experimental data on manganites \cite{HaMaLoKo04}
and also \chem{TiO_2} \cite{Ma90}, and its occurrence in the IC regime points
toward the inadequacy of standard small-polaron theory for such materials.

Finally, for $\Ep/t=4$ (figure~\ref{fig:vd_adiab}(c)), we find a maximum
close to that of the small-polaron result \cite{Em93} which peaks at $2\Ep$
and is included for reference, with the parameter $\overline{\sigma}_0$
chosen so as to get the same $\Sr$. For technical reasons, the optical
conductivity in the SC regime (given analytically by a sum of $\delta$-peaks)
is calculated using an artificial broadening (see many-electron case).

The exact numerical results, however, reveal that the parameters chosen do
not fall into the true small-polaron regime, as indicated by a slight
asymmetry, a maximum clearly below $2\Ep$, and three low-energy peaks (not
related to finite-size effects) which diminish with increasing $\Ep$.  The
latter originate from transitions between the (dispersive) coherent band and
dispersionless phonon satellites (cf figure~9(b) of \cite{LoHoFe06}), and are
absent in the analytical results probably due to an overestimation of
band-narrowing in the adiabatic regime. The possibility of such
overestimation by the Lang-Firsov small-polaron theory in the adiabatic case
was qualitatively illustrated for the two-site model in \cite{AlKaRa94}, and
the analytical results of figure~\ref{fig:vd_adiab}(c) are expected in the SC
(small-polaron) regime.

Figure~\ref{fig:vd_nonadiab} shows $\sir(\om)$ at weak coupling in the
non-adiabatic regime, taking $\om_0/t=4$. As multi-phonon excitations have
small weight due to $g^2\ll1$, the absorption is virtually restricted to the
interval $[\om_0,\om_0+2W]$. Here the analytical WC approach---restricted to
one-phonon excitations, see equation~(\ref{eq:oc_wc})---yields even better
agreement than in figure~\ref{fig:vd_adiab}(a).

Overall, the analytical approach is capable of reproducing the main features
of the optical conductivity in the low-density regime, including the increase
of the frequency range for absorption (figure~\ref{fig:vd_adiab}).
Furthermore, in accordance with figure~\ref{fig:vd_drudesumrule}, $\Sr$ is
small at weak coupling, takes on a maximum in the IC regime, and decreases
again due to the suppression of $\varepsilon_\mathrm{kin}$ (via the sum rule)
at strong coupling.

\begin{figure}
  \centering
  \includegraphics[width=0.5\textwidth]{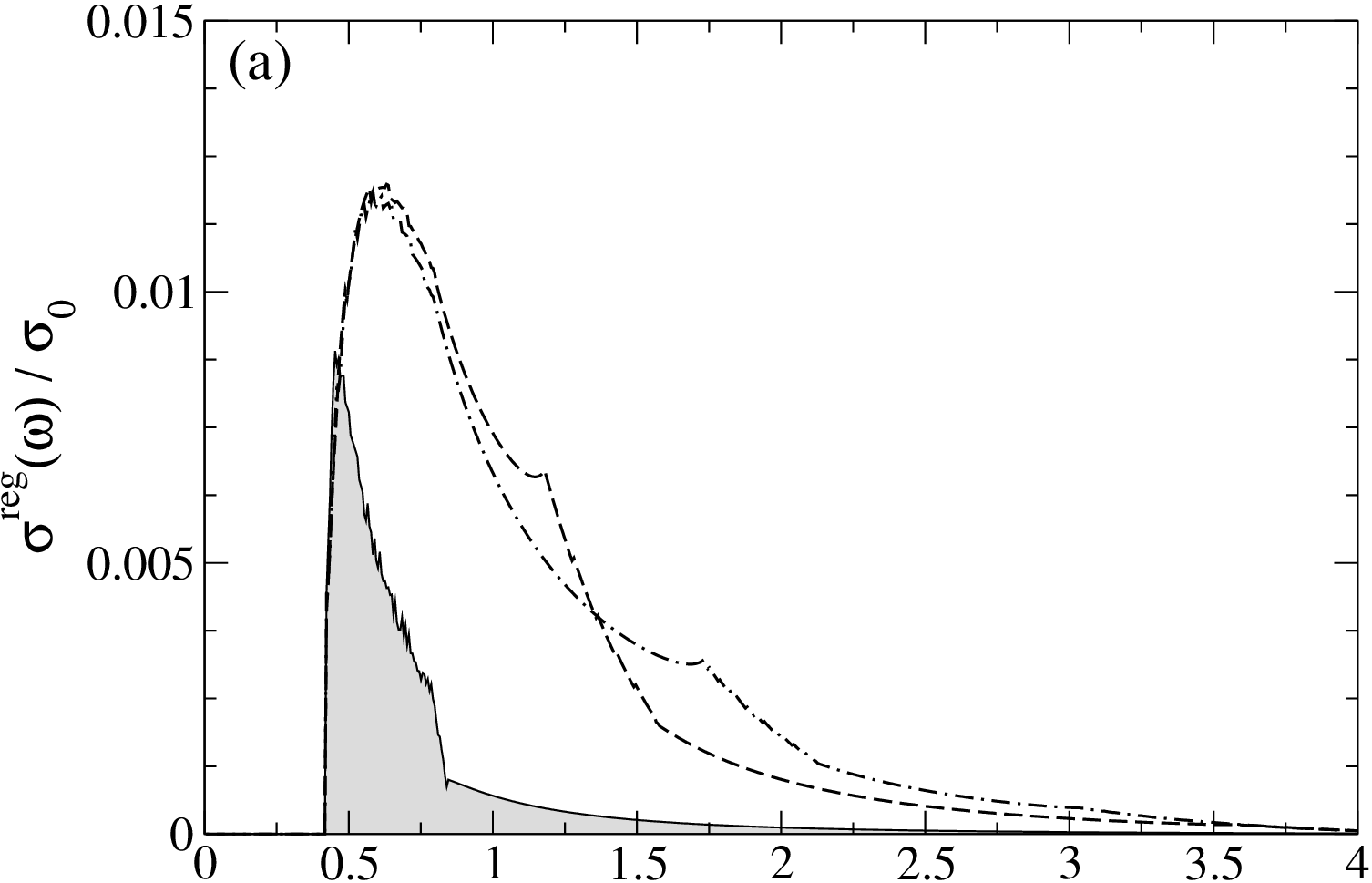}\\
  \includegraphics[width=0.5\textwidth]{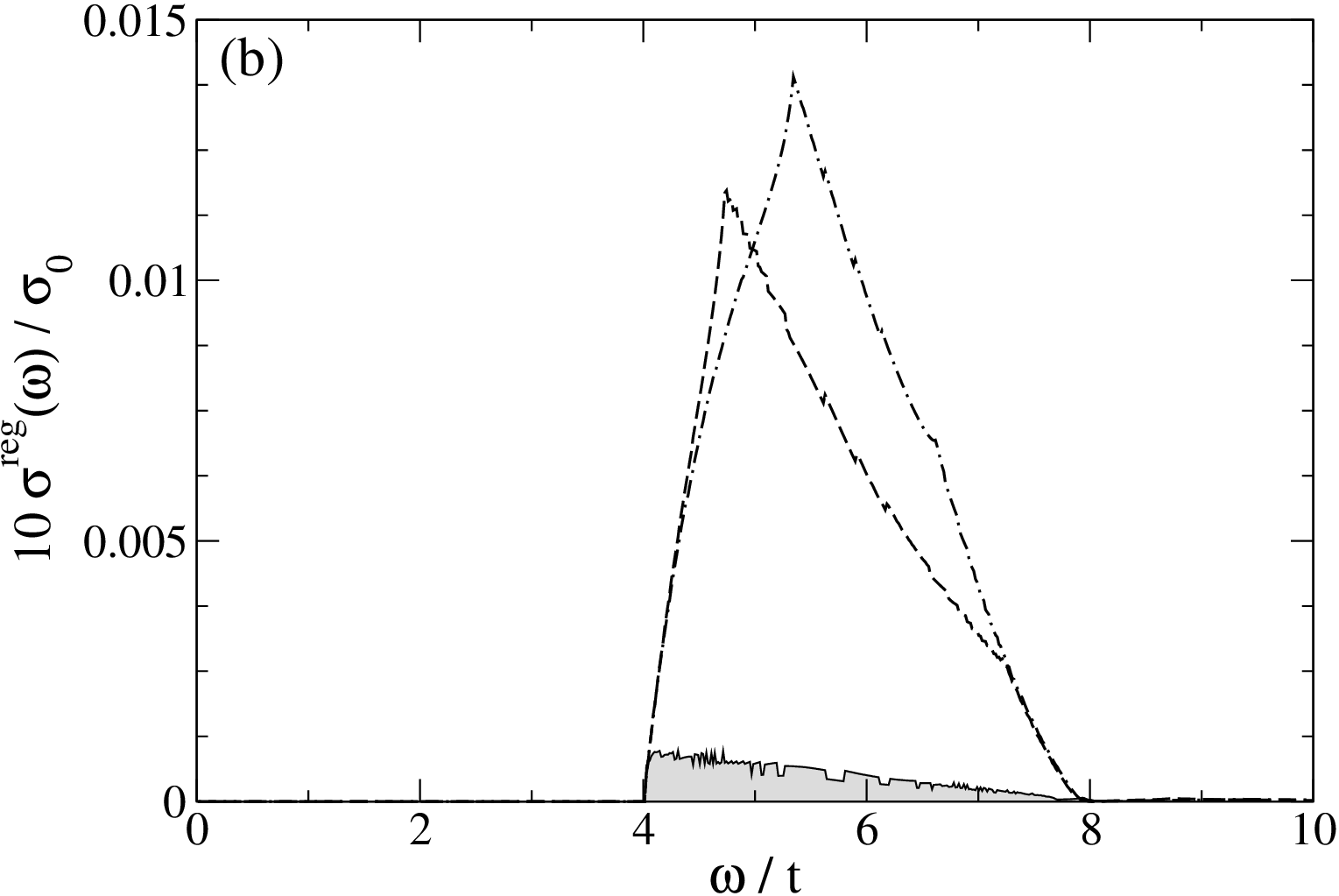}
  \caption{\label{fig:wc_densitydep} Analytical results for $\sir(\om)$ from the WC approximation for
    $n=0.1$ ($\full$), $n=0.3$ ($\dashed$) and $n=0.4$ ($\chain$). Here
    $\Ep/t=0.1$, (a) $\om_0/t=0.4$ and (b) $\om_0/t=4$. The corresponding
    results for $\Sr$ are given in figure~\ref{fig:an_sumrule}.}
\end{figure}

\subsubsection{Many electrons}

There are some issues concerning the analytical results shown here which
deserve attention. First, we pointed out in \cite{LoHoFe06} that the total
spectral weight $w=\int\,\rmd k\int\,\rmd\om A(k,\om)\leq1$ contained in the
spectral function (\ie, its norm) provides a first indication for the quality
of the analytical approximation for a given set of parameters.

The IC approach reproduces the WC (SC) approximation in the WC (SC) regime
\cite{LoHoFe06}. In the vicinity of these limits, it yields results very
similar to the WC/SC approximation, with $w\lesssim 1$. In the true IC regime
($\lambda=\Ep/2t\approx1$ or $g^2=\Ep/\om_0\approx1$), the norm $w$ becomes
as small as 0.7 in the worst case considered (for the parameters of
figure~\ref{fig:ic_densitydep}(a)), with $w$ also depending on the radius
$R$, whose ``optimal'' value is determined by minimising the energy within
the Hartree approach. Despite these deviations, observables such as the
kinetic energy (see table~\ref{tab:Ekin}) or the total energy (see
figure~13(a) in \cite{LoHoFe06}) are in satisfactory agreement with exact
results. Moreover, the optical conductivity---calculated from the spectral
function---is mainly determined by states near the Fermi level, so that
missing high-energy incoherent processes are irrelevant, and $w<1$ does not
necessarily lead to poor results.

For some parameters, the spectral functions obtained from the IC approach
exhibits small regions in the $(k,\om)$ plane where $A(k,\om)<0$
\cite{LoHoFe06}. To calculate the optical conductivity, we have replaced such
values by zero.

Finally, in the SC regime, an artificial ``softening'' proportional to the
energy resolution is imposed on the $\theta$ functions in
equation~(\ref{eq:oc_sc}) which affects the integrated weight $\Sr$.
Nevertheless, a consistent value for $\Sr$, and hence the normalisation of
$\sigma^\mathrm{reg}(\om)$, can be obtained using
equation~(\ref{eq:app:sreg}).

\begin{figure}
  \centering
  \includegraphics[width=0.495\textwidth]{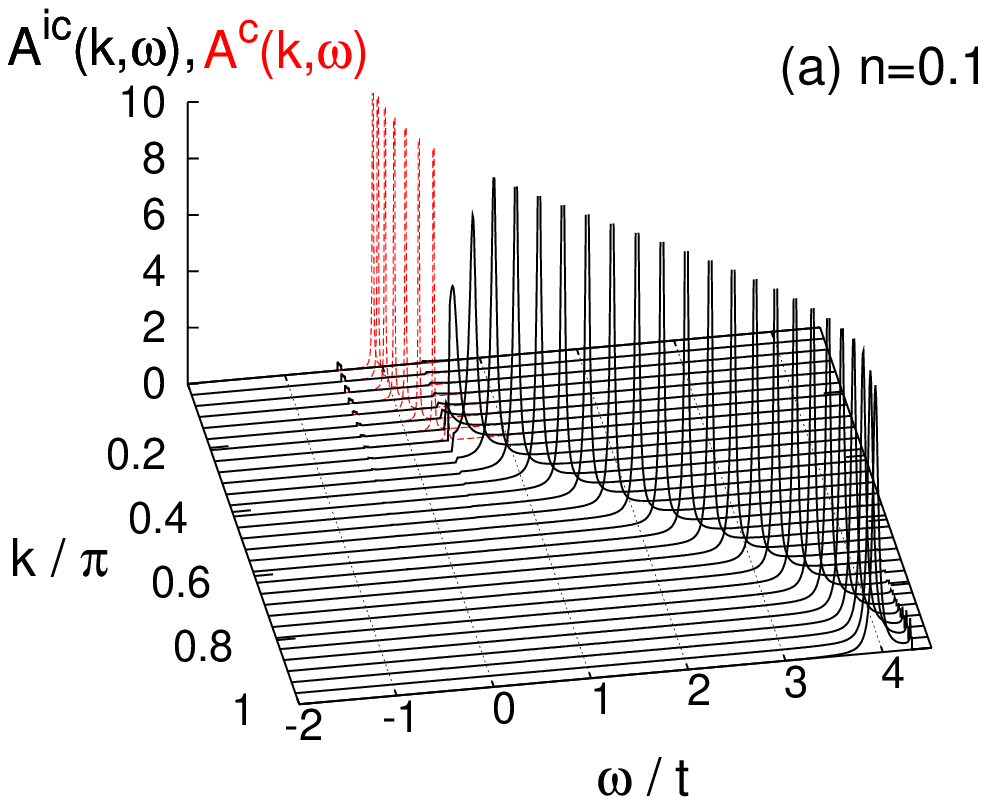}
  \includegraphics[width=0.495\textwidth]{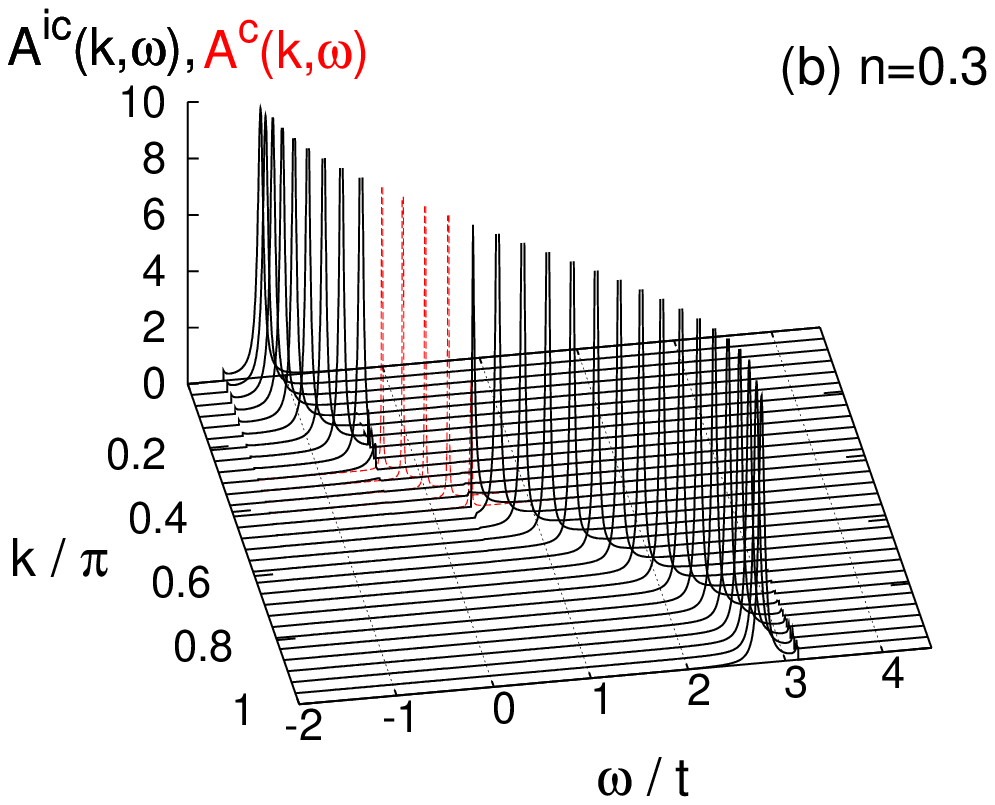}
  \caption{\label{fig:wc_akw}%
    (colour online) Coherent ($A^\mathrm{c}$, $\dashed$) and incoherent
    ($A^\mathrm{ic}$, $\full$) parts of the  WC electronic spectral function 
    for $\om_0/t=0.4$ and $\Ep/t=0.1$
    \cite{LoHoFe06}.}
\end{figure}

\paragraph{Weak coupling}

We begin with the density dependence of $\sir(\om)$.
Figure~\ref{fig:wc_densitydep} shows results for $\Ep/t=0.1$, and we first
consider the adiabatic case $\om_0/t=0.4$ in panel (a). The spectra exhibit
several non-trivial features. Apart from the absorption threshold at $\om_0$,
we find an asymmetric peak exhibiting small but noticeable wiggles which
extend over an interval $\om_0$ centred at the maximum, and diminish with
increasing $n$. 

The origin of this feature are transitions of type II, as can be seen by
separating the contributions I and II to equation~(\ref{eq:oc_wc}), and
looking at the corresponding spectral function $A(k,\om)$ shown in
figure~\ref{fig:wc_akw}. 

Non-zero contributions to the second term in equation~(\ref{eq:oc_wc}) exist
for $\om\in(\om_\mathrm{min},\om_\mathrm{max})$. For a given $\om$ from this
interval, the range of $k$-values contributing to $\sigma^\mathrm{reg}(\om)$
is defined by the condition $A^\mathrm{ic}(k,E_k-\mu-\om)>0$, with
$E_k-\mu>0$. The absorption starts at $\om_\mathrm{min}=\om_0$, with
a single $k$-value $k_\mathrm{min}=k_\mathrm{F}$. The upper cut-off for
the adiabatic case (figure~\ref{fig:wc_densitydep}(a)) is given as
$\om_\mathrm{max}=2\om_0+|-W-\mu|$, corresponding to a single value
$k_\mathrm{max}$ defined by $E_{k_\mathrm{max}}-\mu=\om_0$. In the
antiadiabatic case (figure~\ref{fig:wc_densitydep}(b)), for which the
entire coherent band lies inside the interval $(-\om_0,\om_0)$, 
$\om_\mathrm{max}=\om_0+2W$ corresponding to $k_\mathrm{max}=\pi$ with the
highest energy $W-\mu$.

The shape of the absorption curve in the interval
$(\om_\mathrm{min},\om_\mathrm{max})$ depends on the $k$-interval
contributing to the transition II for fixed $\om$, and the corresponding
values of $A^\mathrm{ic}(k,E_k-\mu-\om)$. For the anti-adiabatic case,
figure~\ref{fig:wc_densitydep}(b) shows sharp maxima at the $n$-dependent
positions $\om_1=\om_0+|-W-\mu|$. This feature may be related to the increase
of the range of contributing $k$-values, beginning with the single value
$k=k_\mathrm{F}$ for $\om=\om_0$, up to the interval $(k_\mathrm{F},k_1)$ for
$\om_1$, with $k_1$ determined by $E_{k_1}-\mu=|-W-\mu|$.

For $n=0.1$, there is very little weight contained in incoherent states below
$\mu$ (figure~\ref{fig:wc_akw}(a)). In fact, there exists only a single,
small and narrow peak with negligible dispersion, and transitions from the
latter to a coherent state can occur for $\om_0\leq\om\leq2\om_0$, \ie, for
final states between $E_k-\mu=0$ and the upper edge of the coherent band
$E_k-\mu=\om_0$. The wiggles, more pronounced for smaller numbers of
$k$-values, result from transitions to the discrete $\delta$-peaks of the
coherent spectrum.

Above $\om=2\om_0$, for $n=0.1$, the transitions of type II fall off quickly
with increasing $\om$ and go to zero for $\om>2\om_0+|-W-\mu|\approx0.87$. In
contrast, for larger $n$, the energy interval of incoherent states below the
Fermi level $|-W-\mu|$ (and the
weight contained in the latter) is significantly larger, yielding the cut-offs of
$\om\approx1.6$ and $\om\approx2.15$. The high-frequency optical response is
much smoother since the incoherent parts of the spectral function are
broadened proportional to $\Ep$, and $\Sr(\om)$ increases in accordance with
figure~\ref{fig:an_sumrule}.

Finally, the long high-energy tail in
$\sir(\om)$ seen in figure~\ref{fig:wc_densitydep}(a) for all $n$ is due to
incoherent excitations of type $D$ (figure~\ref{fig:trans_wc}), which have a
maximum energy of $2(W+\om_0)$ (within the WC approximation). 

\begin{figure}
  \centering
  \includegraphics[width=0.55\textwidth]{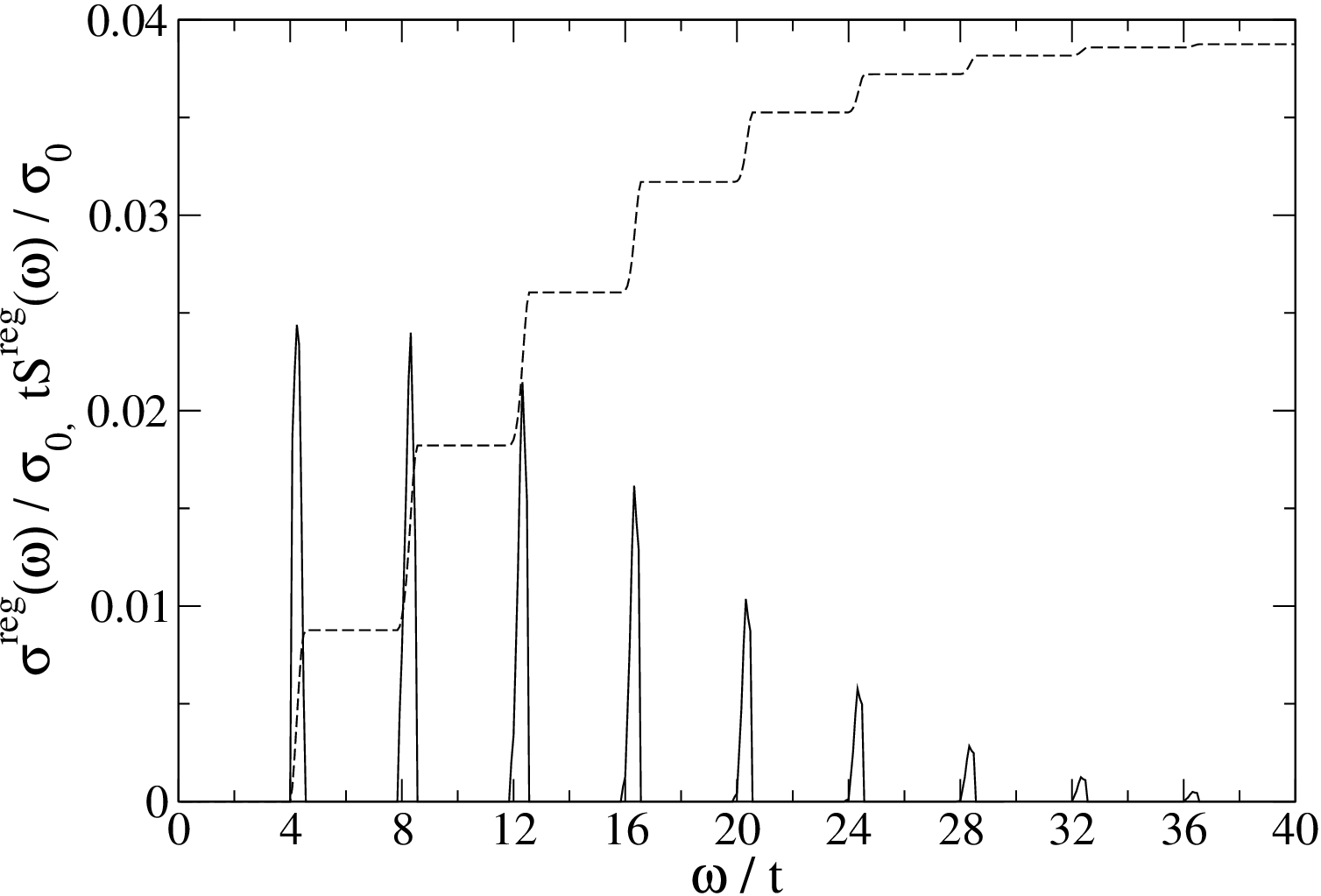}
  \caption{\label{fig:sc_densitydep} Analytical results for $\sir(\om)$
    ($\full$) and $\Sr(\om)$ ($\dashed$) from the SC approximation. Here
    $\om_0/t=4$, $\Ep/t=8$ and $n=0.4$.}
\end{figure}

\begin{figure}
  \centering
  \includegraphics[width=0.55\textwidth]{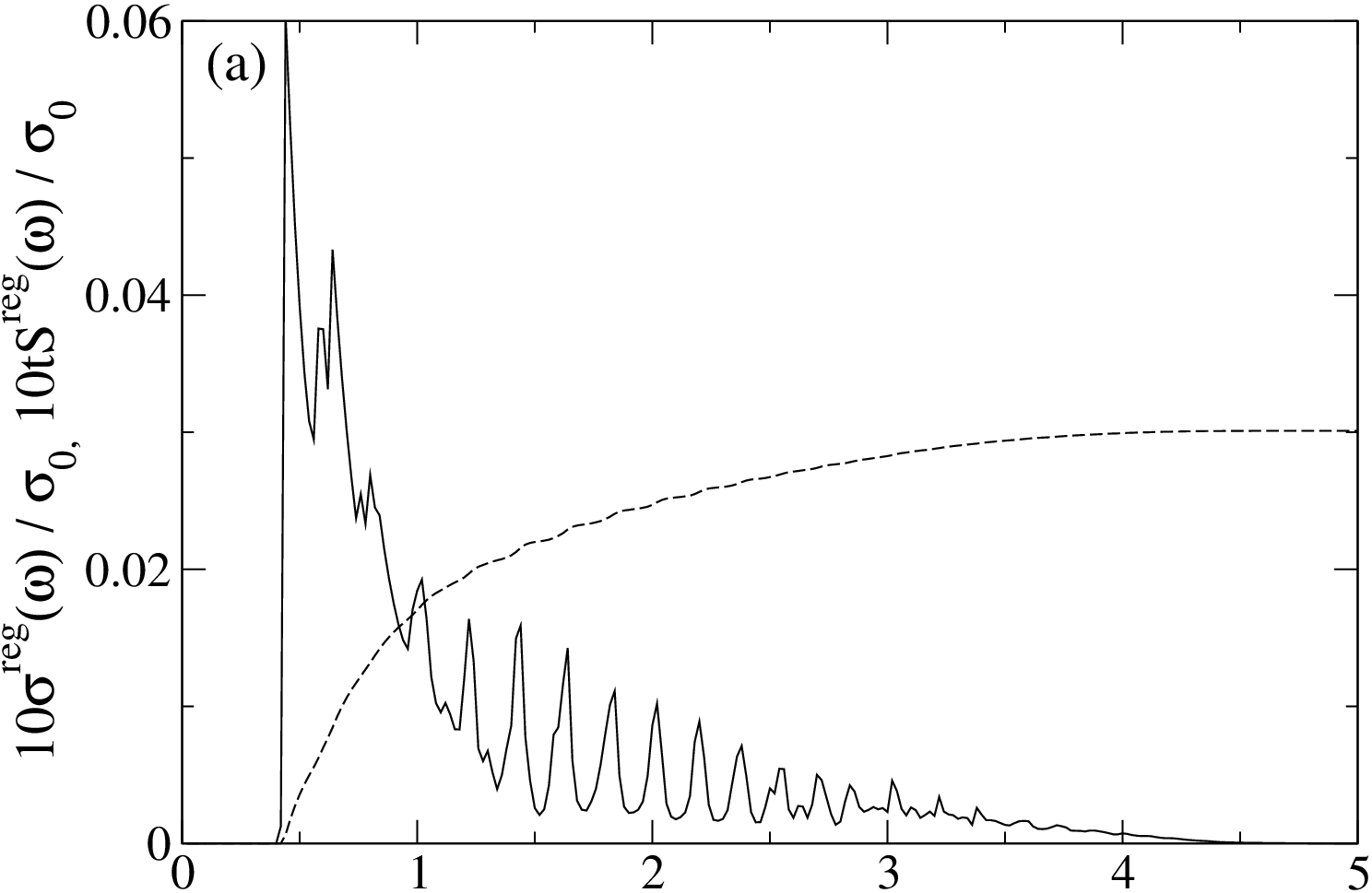}\\
  \includegraphics[width=0.55\textwidth]{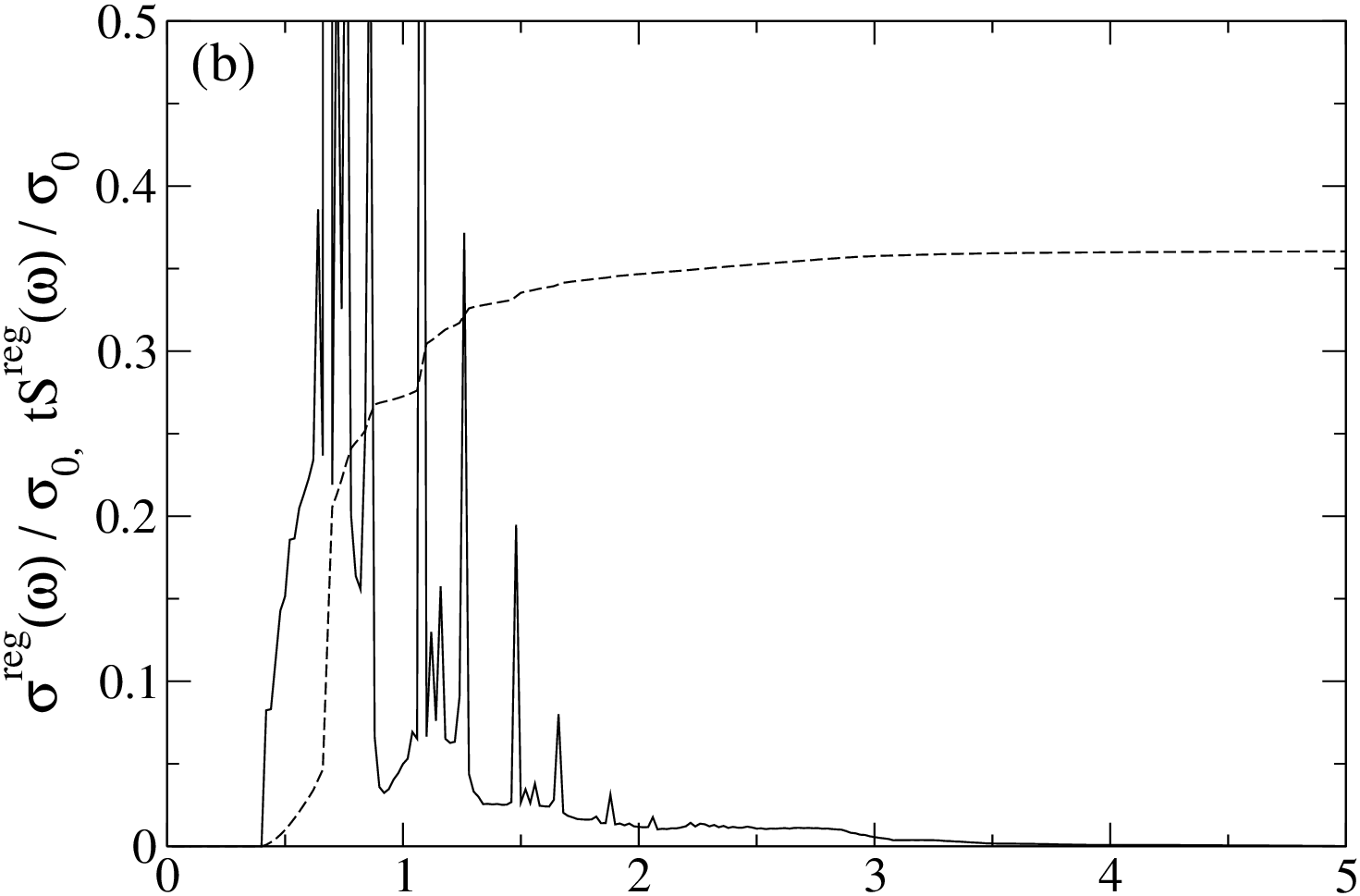}\\
  \includegraphics[width=0.55\textwidth]{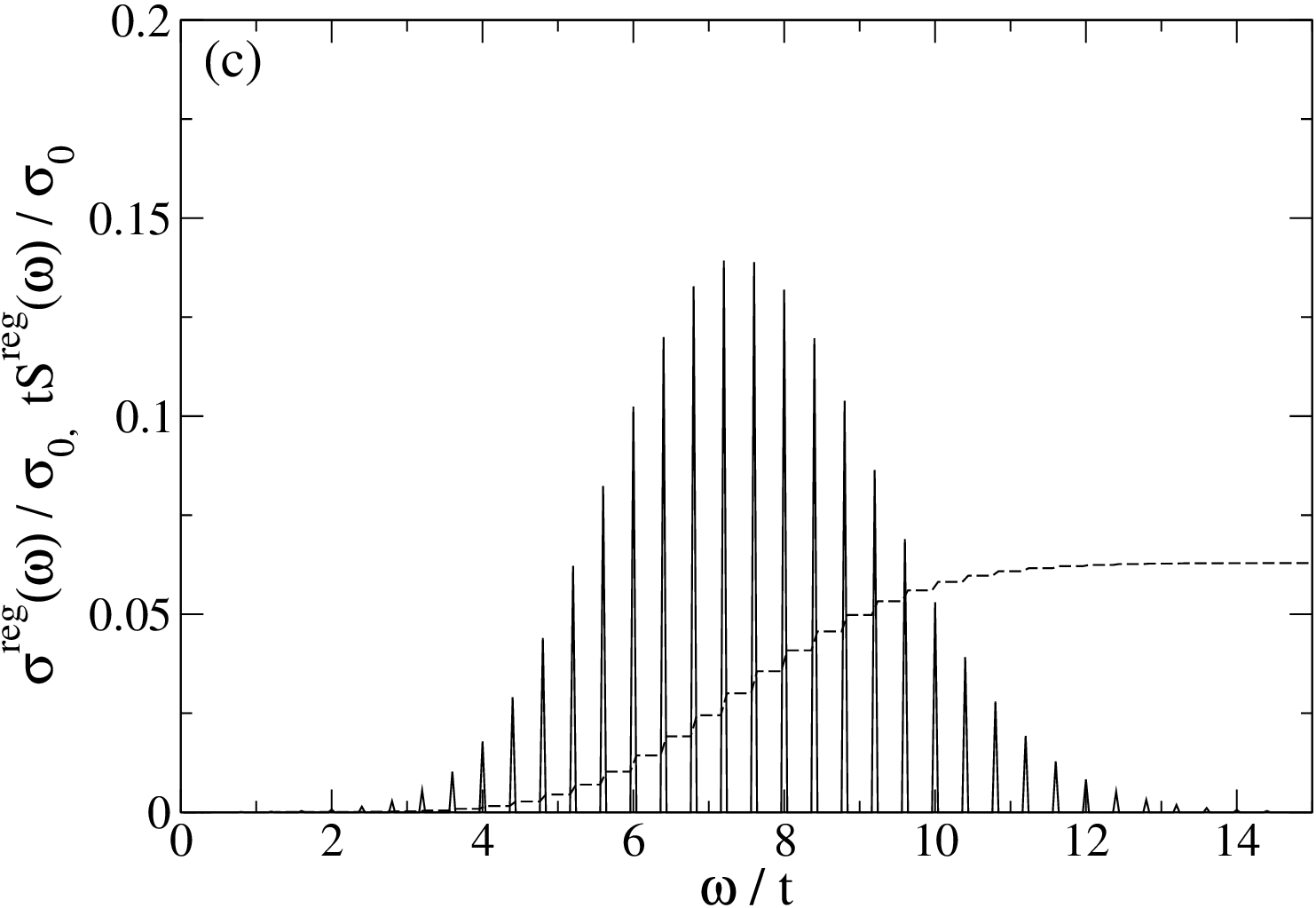}
  \caption{\label{fig:ic_couplingdep} Analytical results for $\sir(\om)$
    ($\full$) and  $\Sr(\om)$ ($\dashed$) from the IC [(a),(b)] respectively
    SC [(c)]    approximation. Here
    $\om_0/t=0.4$, $n=0.4$, and (a) $\Ep/t=0.1$ ($R=1.5$), (b) $\Ep/t=2$
    ($R=1.3$) and (c) $\Ep/t=4$.}
\end{figure}

\paragraph{Strong coupling}

Figure~\ref{fig:trans_ic} illustrates the contributions to
$\sigma^\mathrm{reg}(\om)$ in the SC regime, where the polaron band lies
entirely in the interval $(-\om_0,\om_0)$. Equations~(A1)--(A3) represent
transitions between the states in the polaron band accompanied by
multi-phonon processes. The processes (B) and (C), corresponding to
transitions between the incoherent polaron spectrum and the polaron band
states, again involving multiple phonons, are expected to become important
for intermediate coupling.

Examining the analytical expressions, the most important processes in the
true SC limit ($\lambda,g\gg1$) appear to be those of type (A3), with the
dominating (with respect to $\rme^{-g^2}$) contribution to
$\sigma^\mathrm{reg}(\om)$ given by equation~(\ref{eq:app:resigmaSC}). In
this case, $\sigma^\mathrm{reg}(\om)$ consists of sharp peaks at $\om=s\om_0$
with heights modulated by the Poisson distribution with parameter $2g^2$, as
may be seen in figure~\ref{fig:vd_adiab}(c). The maximum of the conductivity
lies near the maximum of the Poisson distribution determined by
$s=2g^2=2\Ep/\om_0$.

The peak heights in the anti-adiabatic SC case ($g^2=2$) shown in
figure~\ref{fig:sc_densitydep} can be understood by evaluating the relative
weights in front of the integral on the rhs of
equation~(\ref{eq:app:resigmaSC}), yielding $1,1,8/9$ for $s=1,2,3$,
respectively.

The picture of the optical conductivity implied by our findings agrees with
results obtained previously in the framework of small-polaron theory at $T=0$
\cite{Ea84,FeIhLoTrBu94}. The separated peaks may be smeared out by means of
an additional ``smoothing mechanism'' \cite{Ea84} or due to limited
experimental resolution. The evolution of the small-polaron absorption spectrum towards a smooth
function due to damping and finite temperature effects was considered in \cite{LoSt89}, and
Emin \cite{Em93} has presented the absorption spectrum in the form of a
Gaussian centered at $2\Ep$, assuming the broadening of the energy levels of
the localised polaron states.

\paragraph{Crossover from weak to strong coupling}

Figure~\ref{fig:ic_couplingdep} illustrates the crossover from
weak to intermediate to strong coupling for a fixed density $n=0.4$ in the
adiabatic regime $\om_0/t=0.4$ within the IC/SC approach. The
corresponding spectral functions can be found in figure~14 of
\cite{LoHoFe06}, and we use the same values of the parameter $R$ which
decreases with increasing $\Ep$ reflecting a decrease of the polaron size.

The position of the maximum in $\sir(\om)$ changes
significantly as a function of $\Ep$. For weak coupling, it is located just
above the absorption threshold at $\om=\om_0$
(figure~\ref{fig:ic_couplingdep}(a)), whereas for $\Ep/t=2$ it lies near
$1.5\om_0$.  Nevertheless, the frequency range for absorption in panels (a)
and (b) is very similar. For $\Ep/t=4$, similar to the low-density case shown
in figure~\ref{fig:vd_adiab}(c), $\sir(\om)$ peaks near the small-polaron
value $2\Ep$. The analytical approach overestimates the band narrowing,
yielding SC behaviour even for a value $\Ep/t=4$ lying at the border to the
SC region. Since for this value of $\Ep$ the IC and SC approximations yield
virtually identical results, we have used the SC approximation to obtain
a consistent value of $\Sr$. The similarity of the SC results for $n=0.1$ in
figure~\ref{fig:vd_adiab}(c) and $n=0.4$ in figure~\ref{fig:ic_couplingdep}(c) is
due to the rather weak density dependence of our analytical results pointed
out before in the case of the spectral function \cite{LoHoFe06}.

The evolution of $\Sr$ with increasing $\Ep$ is similar to that observed in
the one-electron case (figure~\ref{fig:vd_drudesumrule}(a)), with a maximum
in the IC regime. Whereas for weak coupling transport is mainly coherent (as
reflected by a large Drude weight $t\mathcal{D}/\sigma_0\approx0.61\gg
tS^\mathrm{reg}/\sigma_0$), $\Sr$ increases noticeably for $\Ep/t=2$ owing to
enhanced incoherent scattering. For strong coupling, since $\mathcal{D}\to0$,
the reduced kinetic energy suppresses $\Sr$ via the f-sum rule.

\begin{figure}
  \centering
  \includegraphics[height=0.35\textwidth]{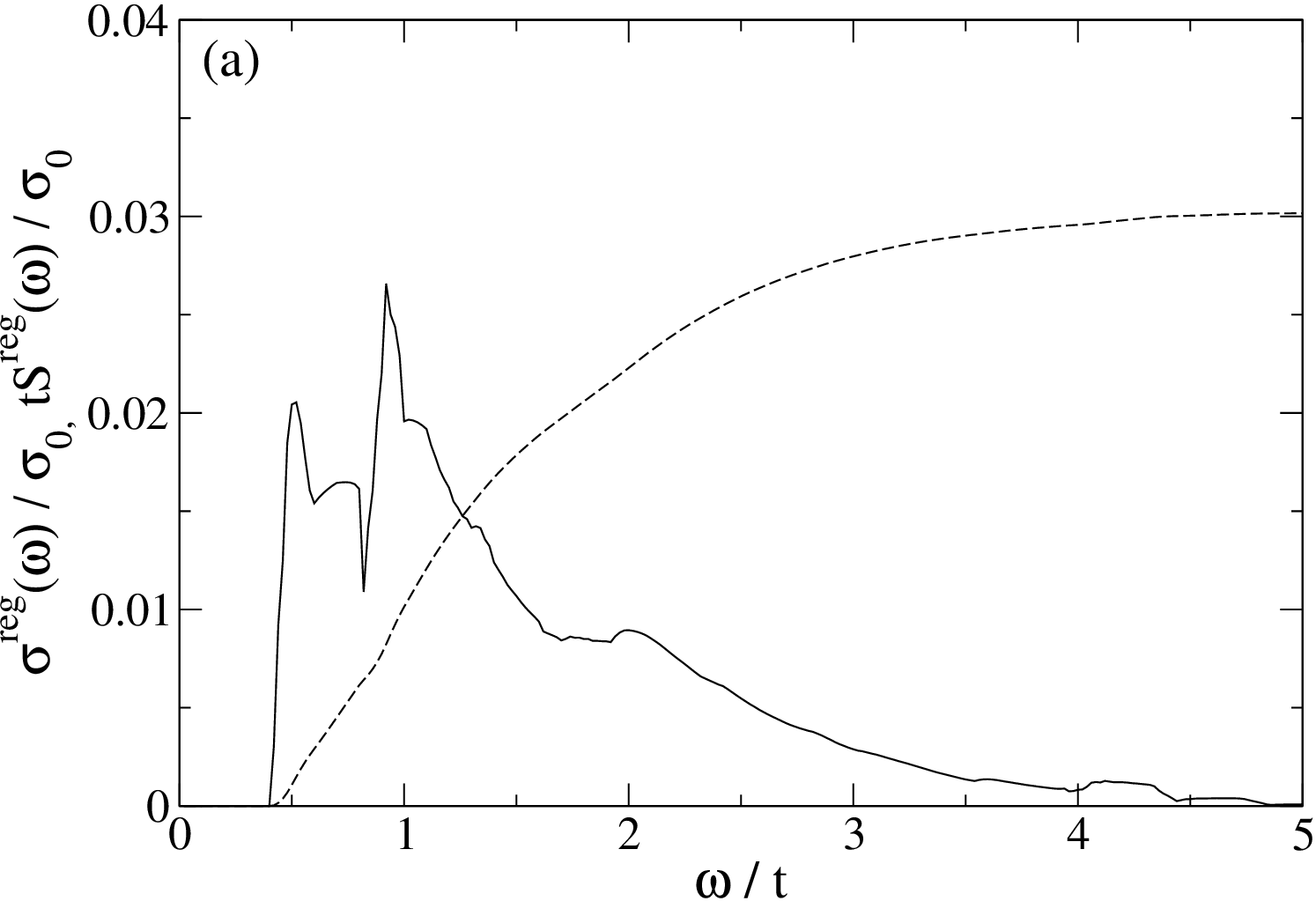}
  \includegraphics[height=0.35\textwidth]{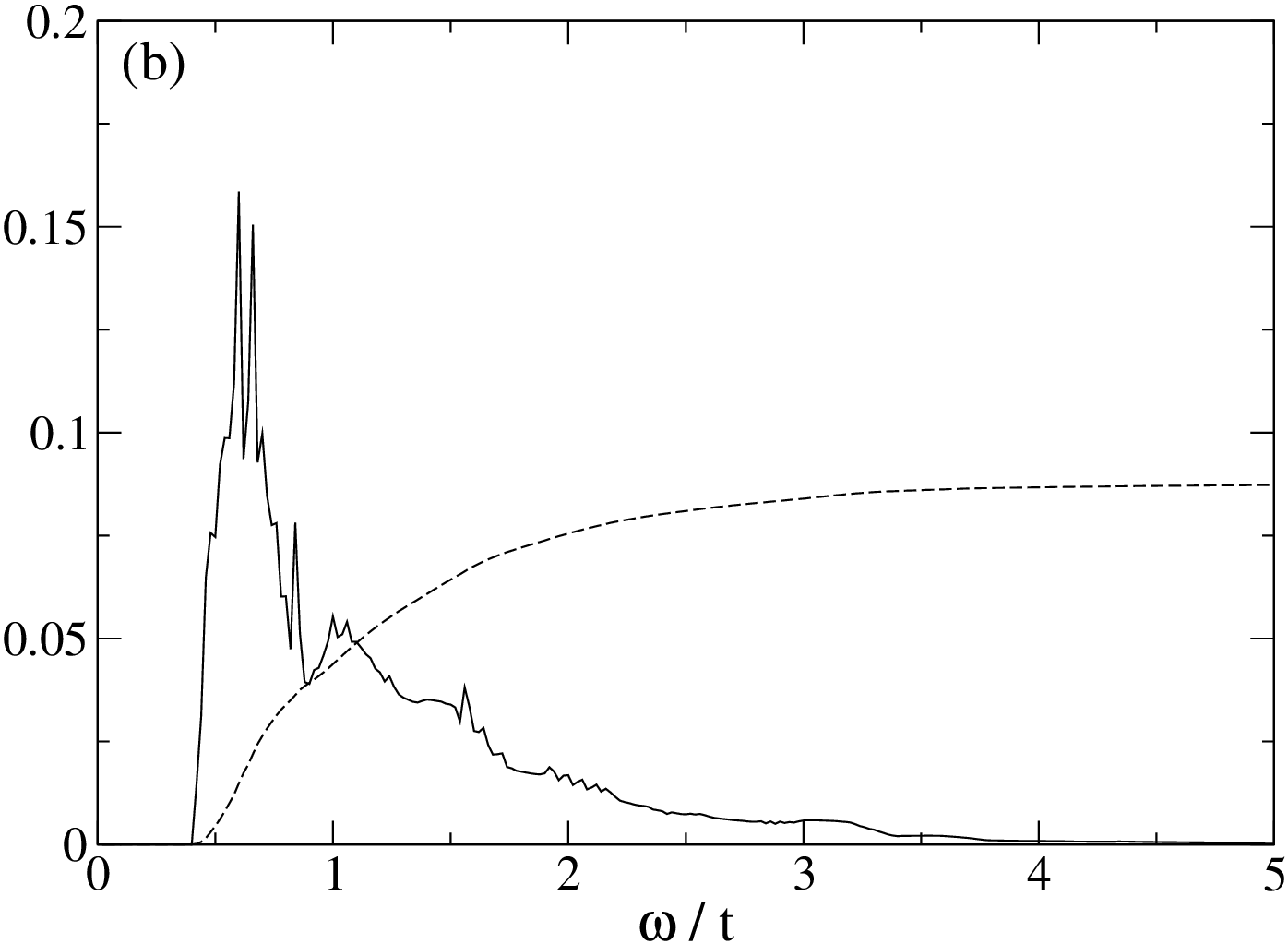}
  \caption{\label{fig:ic_densitydep}  
  Analytical results for $\sir(\om)$ ($\full$) and 
  $\Sr(\om)$ ($\dashed$) from the IC approximation. Here $\om_0/t=0.4$,
  $\Ep/t=2.0$, (a) $n=0.1$ ($R=2.1$) and (b) $n=0.3$ ($R=1.5$).}
\end{figure}

\paragraph{Density dependence at intermediate coupling}

We finally examine the density dependence of the optical response in the IC
regime.  To make a connection with previous work \cite{WeBiHoScFe05}, we take
$\om_0/t=0.4$ and $\Ep/t=2$. Figure~\ref{fig:ic_densitydep} depicts results
for $\sir(\om)$ for $n=0.1$ and $n=0.3$, whereas the case $n=0.4$ is reported
in figure~\ref{fig:ic_couplingdep}(b).

The low-density case $n=0.1$ has already been shown in
figure~\ref{fig:vd_adiab}(b), and results are presented without rescaling in
figure~\ref{fig:ic_densitydep}(a) for reference. Compared to $n=0.3$
(figure~\ref{fig:ic_densitydep}(b)), we observe a transfer of spectral weight
from high to low frequencies, causing a strong reduction of the broad
high-energy hump. This trend continues upon increasing $n$ even further, as
can be seen from figure~\ref{fig:ic_couplingdep}(b). Besides, the continuous
increase of $\Sr$ with increasing $n$ is similar to the WC case depicted in
figure~\ref{fig:an_sumrule}(a).

These features of the optical conductivity suggest that the increase of the
kinetic energy with $n$, as shown in table~\ref{tab:Ekin}, has a more
profound origin than the mere increase of the charge carrier density. In this
connection, it is necessary to discuss the dependence of the ``optimal''
polaron radius $R$ on density $n$. The values reported in the caption of
figures~\ref{fig:ic_couplingdep} and~\ref{fig:ic_densitydep} illustrate that
within our variational approach, $R$ decreases with increasing carrier
density, \ie, $R=2.1$ for $n=0.1$, $R=1.5$ for $n=0.3$, and $R=1.3$ for
$n=0.4$. This seems to be at odds with the crossover from polaronic behaviour
at small $n$ to metal-like behaviour at larger $n$ observed numerically
\cite{HoNevdLWeLoFe04,HoHaWeFe06}, since a decrease of $R$ with increasing
$\Ep$ at fixed $n$ generally enhances the polaronic character of the spectra
\cite{LoHoFe06}.

Whereas the dependence of $\varepsilon_\mathrm{kin}$ on the coupling strength
at fixed $n$ may be interpreted as the crossover between the small-polaron
and large-polaron regime, the concentration dependence of
$\varepsilon_\mathrm{kin}$ seems to be more involved
\cite{HoNevdLWeLoFe04,HoHaWeFe06}. Indeed, according to the variational
treatment of the IC case, the concentration dependence of the kinetic energy
and optical conductivity is to be explained by the $n$-dependence of the
parameters $\gamma,\gammab,\eta$ defined by
equations~(\ref{eq:trans_ic}),(\ref{eq:gammab}). These parameters are
determined, using the variational variables $R$, by the energy balance of the
Hartree energy of polarons and the energy of the lattice deformation
background \cite{LoHoFe06}, which at $\lambda=1$ may be quite significant.
Although the optimal $R$ increases with decreasing $n$, this energy balance
leads to an increase of kinetic energy with increasing $n$, as the resulting
increase of the lattice deformation is sufficient to enhance the mobility of
charge carriers.

This picture may be supplemented and improved by a more detailed insight
provided by the electron-lattice and electron-electron correlations
calculated numerically in \cite{HoHaWeFe06}. According to these results, the
increase of the carrier concentration is accompanied by the short-range
development of the charge-density wave, which is connected with the spreading
of the inhomogeneous lattice deformation about the centre of the polaron.
Consequently, the analytical calculations based on the Hartree
energy---taking into account only the averaged energy of the lattice
deformation background---cannot describe these features revealed by numerical
calculations. Nevertheless, the optical spectra turn out to reproduce the
main features found numerically \cite{WeBiHoScFe05,HoHaWeFe06}, and the
dependence of the total energy on $n$ is in good agreement with exact data
\cite{HoHaWeFe06}.

\section{Discussion and conclusions}\label{sec:conclusion}

The results of section~\ref{sec:results:opt} suggest that the analytical
approach captures the main features of polaronic systems with finite carrier
density. Therefore, it is interesting to relate our findings to the polaronic
excitations observed in manganites, for which no general theory is available.
To be more specific, we consider results for thin films of
\chem{La_{2/3}Sr_{1/3}MnO_3} (LSMO) and \chem{La_{2/3}Ca_{1/3}MnO_3} (LCMO)
containing a characteristic mid-infrared peak usually assigned to polarons
\cite{HaMaLoKo04}. The shape and position of this peak differs substantially
for the two materials, and has been interpreted in terms of large and small
polarons, respectively \cite{PhysRevLett.81.1517,HaMaLoKo04}. Apart from the
shift of the maximum in $\sigma^\mathrm{reg}(\om)$ to higher energies when
going from LSMO to LCMO, the peak is highly asymmetric in LSMO---with a sharp
onset just below the maximum and a high-energy tail---and much more symmetric
in LCMO.

In small-polaron theory, the maximum in $\sigma^\mathrm{reg}(\om)$ occurs at
$2\Ep$ \cite{Ma90}, so that experiments can in principle be used to determine
$\Ep$. However, for small-polaron theory to be applicable, the resulting
value has to be consistent with the high-temperature activation energy of the
dc conductivity, \ie, $4E_\mathrm{a}=\Ep$ \cite{Ma90}. Experimentally, it
turns out that this is usually not the case, indicating that small-polaron
theory is invalid. This problem has been pointed out in
\cite{HoEd01,David_AiP,HaMaLoKo04}, and a many-electron approach valid at
intermediate coupling was proposed \cite{Gr01,HoEd01}. The basic assumptions
are that perovskite LSMO falls into the WC regime, whereas LCMO is
characterised by intermediate to strong coupling, in accordance with the
Curie temperatures $T_\mathrm{C}\approx370$K respectively
$T_\mathrm{C}\approx270$K \cite{Gr01}, and that the polaronic character of
quasiparticles is enhanced in thin films \cite{HoEd01}. Hartinger \etal
\cite{HaMaLoKo04} compare various theories to their data, concluding that
none is able to satisfactorily explain the lineshape of the polaron peak.

Based on the results of our simple analytical theory, we can make the
following remarks. The change of shape from asymmetric to symmetric and the
position of the maximum is observed when comparing the WC and IC/SC regimes.
In particular, the WC approach features a sharp onset and a long high-energy
tail, whereas the IC/SC results are reminiscent of the Poisson distribution
of phonons. These main features can already be observed in our exact
one-electron results in figure~\ref{fig:vd_adiab}. Apart from setting the
order of magnitude of $\sigma^\mathrm{reg}(\om)$, the many-polaron effects
discussed above are of minor importance for weak coupling (as appropriate for
LSMO), but renormalise polaronic features toward ``metallic'' (or more
WC-like) behaviour in the IC regime \cite{HoNevdLWeLoFe04}. Although polaron
effects are enhanced in thin films as used in \cite{HaMaLoKo04}, the
discrepancy between $\Ep$ as deduced from $\sigma^\mathrm{reg}(\om)$ and
$E_\mathrm{a}$ for LCMO in \cite{HaMaLoKo04} suggests that the sample lies
near ($T_\mathrm{C}\approx240$K) but not in the SC regime. The true SC case
seems to be realised in the layered manganite \chem{La_{1.2}Sr_{1.8}Mn_2O_7},
for which angular-resolved photoemission spectra \cite{De98} suggest the
existence of small polarons even in the ferromagnetic state
\cite{AlBr99_2,HoEd01}.

The present approach correctly predicts coherent states at the
Fermi level giving rise to a Drude response, and reproduces standard
small-polaron results in the SC limit. This is in contrast to the analytical
treatment of the Holstein double-exchange model in \cite{Gr01}, which
includes the coupling of local and itinerant spins in the manganites. The WC
theory for a gas of polarons proposed in \cite{TeDe01} may in principle explain the optical conductivity of LSMO, but
it is not clear why long-range electron-phonon
interaction is not screened in a dense metallic system. Exact results for the
Fr\"ohlich model with one electron have been given in \cite{MiNaPrSaSv03}. A
model of optical absorption based on transitions between ground and excited
mixed polaron states, formed by the hybridisation of the narrow small-polaron
band, and the wide large-polaron band was presented in \cite{Ea84,Ea84_2}.
However, this model is dealing with a number of constants, not explicitly
connected with the parameters of the Holstein Hamiltonian, which are to be
determined by fitting the experimental absorption curves. Besides, it is not
clear how the crossover from the SC to the WC case and the optical absorption
in the WC regime can be described in this framework.

To conclude, we have developed an analytical theory for the optical
conductivity of finite-density polaronic systems, based on previous work on
the single-particle spectral function. The spectra are surprisingly rich and
in satisfactory agreement with exact results. Together with previous
numerical results \cite{WeBiHoScFe05}, the present work provides a better
understanding of the optical conductivity of dense systems with large or
small polarons. Finally, it can explain some of the generic polaronic
features observed in the manganites, but our electron-phonon model is
too simple to make quantitative predictions.


\ack

We gratefully acknowledge financial support by the DFG and the Academy of Sciences
of the Czech Republic (ASCR) under Grant Nr.~436 TSE 113/33/0-3, and the Austrian
Science Fund (FWF) through the Erwin-Schr\"odinger Grant No~J2583. We thank
G~Wellein for producing the ED data for table~\ref{tab:Ekin}.

\appendix

\section{}\label{sec:a}

The fermion spectral function (of electrons or polarons) can be written as
\begin{equation}\label{eq:A1}
  A(k,\omega)
  =
  -\frac{1}{\pi}
  \frac{\mathrm{Im}\,\Sigma(k,\omega)}
  {[\omega-(\xi_k+\eta) - \mathrm{Re}\,\Sigma(k,\omega)]^2 +
    [\mathrm{Im}\,\Sigma(k,\omega)]^2}
  \,,
\end{equation}
with the fermion self-energy $\Sigma(k,\om)$, and the free band dispersion
of electrons (polarons) $\xi_k=-W\cos{k}$ ($\xi_k=-\tilde{W}\cos{k}$).

For the Hamiltonian~(\ref{eq:HM}), the coherent part of the spectrum,
$A^\mathrm{c}(k,\om)$---characterised by $\Im\,\Sigma(k,\om)=0$---is non-zero
for $|\om|<\om_0$. In this frequency interval, 
\begin{equation}\label{eq:sc_coherent}
  A^\mathrm{c}(k,\om)
  =
  z_k \delta[\om - (E_k + \eta)]
  \,,
\end{equation}
where $E_k$ is the solution of the equation
\begin{equation}\label{eq:sc_band}
  E_k 
  =
  \xi_k + \Re\,\Sigma(k,E_k+\eta)
  \,,
\end{equation}
and
\begin{equation}\label{eq:zk}
  z_k^{-1}
  =
  \left|
   1  - [\partial_\omega\Re\,\Sigma(k,\omega)]_{\omega=E_k+\eta}
  \right|
  \,.
\end{equation}
Accordingly, the incoherent part of the spectrum, denoted as
$A^\mathrm{ic}(k,\om)$, is non-zero for $|\om|>\om_0$ and can be calculated
according to equation~(\ref{eq:A1}).

The optical conductivity is determined by the electronic spectral
function $\Ae(k,\om)$. However, in the SC case, the correct fermionic
quasiparticles are small polarons, and $\Ae(k,\om)$ has to be 
expressed in terms of the polaronic spectral function $\Ap(k,\om)$ via
\cite{LoHoFe06}
\begin{eqnarray}\label{eq:sc_el_spectrum}
  \fl
  \Ae(k,\om)
  =
  \rme^{-g^2}\Ap(k,\om) 
  + \rme^{-g^2}\frac{1}{N}
  \sum_{s\geq1}\frac{(g^2)^s}{s!}
  \nonumber\\
  \hspace*{-1.5em}
  \times
  \sum_{k'}\left[
    \Ap(k',\om-s\om_0) 
    \theta(\om-s\om_0)
    +
    \Ap(k',\om+s\om_0)
    \theta(-\om-s\om_0)
  \right]
\,.
\end{eqnarray}

For the numerical evaluation of equation~(\ref{eq:resigma}), we define
\begin{eqnarray}\label{eq:Ae<}\fl\nonumber
  \Ae^<(k,\om<0)
  =
  \rme^{-g^2} z_k\delta[\om-(\Ek+\eta)]\theta(\om_0+\omega)
  + \rme^{-g^2}
  \sum_{s\geq1}\frac{g^{2s}}{s!}
  \frac{1}{N}
  \\
  \times
  \sum_{k'}z_{k'}
  \delta[\om+s\om_0-(E_{k'}+\eta)]\theta(-\om-s\om_0)\theta[\om+(s+1)\om_0]
\end{eqnarray}
and
\begin{eqnarray}\label{eq:Ae>}\fl\nonumber
  \Ae^>(k,\om>0)
  =
  \rme^{-g^2} z_k 
  \delta[\om-(\Ek+\eta)]\theta(\om_0-\om)
  + \rme^{-g^2}
  \sum_{s\geq1}\frac{g^{2s}}{s!}\frac{1}{N}
  \\
  \times
  \sum_{k'}z_{k'}
  \delta[\om-s\om_0-(E_{k'}+\eta)]
  \theta(\om-s\om_0)\theta[(s+1)\om_0-\om]
  \,,
\end{eqnarray}
as well as
\begin{eqnarray}\nonumber\label{eq:app:aeic}
  \Ae^\mathrm{ic}(k,\om\gtrless0)
  &=&
  \rme^{-g^2} \Ap^\mathrm{ic}(k,\om)\theta(\pm\om-\om_0)
  + \rme^{-g^2}
  \sum_{s\geq1}\frac{g^{2s}}{s!}\frac{1}{N}
  \\
  &&\times
  \sum_{k'}
  \Ap^\mathrm{ic}(k',\om\mp s\om_0)\theta[\pm\om-(s+1)\om_0]
  \,.
\end{eqnarray}
Equations~(\ref{eq:sc_el_spectrum})--(\ref{eq:app:aeic}) for the SC case are
formally the same in the IC case, with $g$ replaced by $\gammab g$ throughout.

The contributions to the integral in equation~(\ref{eq:ABCD}) are then given by
\begin{eqnarray}\nonumber\label{eq:abcddef}
  A 
  &=& 
  \int_{-\infty}^0 \rmd \om' \theta(\om+\om')
  \Ae^<(k,\om')\Ae^>(k,\om'+\om)\,,
  \\\nonumber
  B 
  &=&
  \int_{-\infty}^0 \rmd \om' \theta(\om+\om')
  \Ae^<(k,\om')\Ae^\mathrm{ic}(k,\om'+\om)\,,
  \\\nonumber
  C 
  &=&
  \int_{-\infty}^0 \rmd \om' \theta(\om+\om')
  \Ae^\mathrm{ic}(k,\om')\Ae^>(k,\om'+\om)\,,
  \\  
  D 
  &=&
  \int_{-\infty}^0 \rmd \om' \theta(\om+\om')
  \Ae^\mathrm{ic}(k,\om')\Ae^\mathrm{ic}(k,\om'+\om)
  \,.
\end{eqnarray}

\section{}\label{sec:b}

The Drude formula for the low-frequency conductivity,
\begin{equation}
  \Re\,\sigma(\om)
  =
  \frac{\mathcal{D}}{\pi}\frac{\tau}{1+(\om\tau)^2}
  \,,
\end{equation}
gives at $\om=0$ for the dc conductivity
\begin{equation}\label{eq:app_drude1}
  \sigma(0)
  =
  \mathcal{D}\tau/\pi
\,.
\end{equation}
Here $\tau$ denotes the relaxation time. In the limit $1/\tau\to0$, we obtain
\begin{equation}\label{eq:app_drude}
  \Re\,\sigma(\om)
  =
  \mathcal{D}\delta(\om)
  \,.
\end{equation}

The standard way to deduce the conductivity at $\om=0$ using the Kubo formula
is to take the limit $\om\to0$ of $\Re\,\sigma(\om)$ calculated for $\om>0$
\cite{Ma90}. However, equation~(\ref{eq:resigma}) yields zero for
$0<\om<\om_0$. To derive the Drude-like singularity~(\ref{eq:app_drude}) for
the formulation of the f-sum rule, we shall start with a low but non-zero
temperature $T$. Moreover, the spectral function in
equation~(\ref{eq:resigma}) is assumed to have a finite width---caused by a
very weak additional scattering mechanism---leading to
\begin{equation}\label{eq:app_A}
  A(k,\om')
  =
  \frac{\kappa z_k}{\pi}\frac{\Delta}{[\om'-(\Ek+\eta)]^2+\Delta^2}
\,,
\end{equation}
where $\Delta=(2\tau)^{-1}$. In the WC approach, $\kappa=1$, $\eta=-\mu$, whereas in
the IC/SC approach $\kappa=\exp(-g^2\gammab^2)$, and $\eta$ is given by
equations~(\ref{eq:scC}) respectively~(\ref{eq:trans_ic}).

After substitution of equation~(\ref{eq:img}) into
equation~(\ref{eq:resigma1}), the limit $\om\to0$ gives
\begin{equation}\fl\label{eq:app_si1}
  \lim_{\om\to0}\Re\,\sigma(\om)
  =
  \sigma(0)
  =
  -4\sigma_0\frac{1}{N}\sum_k (\sin k)^2 
  \int_{-\infty}^\infty \rmd\om' [A(k,\om')]^2 \rmd_{\om'} f(\om')
  \,.
\end{equation}
The subsequent limit $T\to0$ is taken using $[\rmd_{\om'} f(\om')]_{T=0}=-\delta(\om')$, which after
substitution of equation~(\ref{eq:app_A}) into equation~(\ref{eq:app_si1})
yields
\begin{equation}
  \lim_{T\to0}\sigma(0)
  =
  \frac{4\kappa^2\sigma_0}{\pi^2}
  \frac{1}{N}
  \sum_k (\sin k)^2 z_k^2 \frac{\Delta^2}{[(\Ek+\eta)^2+\Delta^2]^2}
  \,.
\end{equation}
Converting the sum over $k$ into an integral over $E_k$ we obtain
\begin{equation}
  \sigma(0)
  =
  \frac{4\kappa^2\sigma_0}{\pi^3}
  \int_{\{E_k\}} \rmd E_k (\sin k)^2 \frac{z_k^2}{\partial_k E_k}
  \frac{\Delta^2}{[(\Ek+\eta)^2+\Delta^2]^2}
  \,,
\end{equation}
where the integration is over the energy range of the coherent part of the
spectrum. However, considering $\Delta\to0$, the integrated function is
non-zero only in a narrow  interval around $\Ek+\eta=0$, \ie, for $k\approx
k_\mathrm{F}$. Assuming the slowly varying part of the integrand to be
constant, the integral reduces to
\begin{equation}
  \int_{-\infty}^\infty\rmd E_k
  \frac{\Delta^2}{[(\Ek+\eta)^2+\Delta^2]^2}
  =
  \frac{\pi}{2\Delta} = \pi\tau
  \,.
\end{equation}
Comparison with equation~(\ref{eq:app_drude1}) then yields
\begin{equation}
  \sigma(0)
  =
  \frac{4\kappa^2\sigma_0}{\pi^2}(\sin k_\mathrm{F})^2
  \left[\frac{z^2_k}{\partial_k E_k}\right]_{k=k_\mathrm{F}} \tau
  \equiv
  \mathcal{D}\tau/\pi
  \,.
\end{equation}
According to equation~(\ref{eq:app_drude}), for $1/\tau\to0$, the Drude
singularity takes the form
\begin{equation}\label{eq:app:drude_gen}
  [\Re\,\sigma(\om)]_{D}
  =
  \mathcal{D}\delta(\om)
  =
  \frac{4\kappa^2\sigma_0}{\pi}(\sin k_\mathrm{F})^2
  \left[\frac{z^2_k}{\partial_k E_k}\right]_{k=k_\mathrm{F}} \delta(\om)
  \,.
\end{equation}
In particular, in the WC approximation with $\gammab=0$ in
equation~(\ref{eq:zkEk}) and $\kappa=1$,
\begin{equation}\label{eq:app:drude_wc}
  \mathcal{D}
  =
  2te^2 a z_k\sin k_\mathrm{F}
  \,.
\end{equation}
To check the above considerations, we briefly examine the free-electron case.
In the latter, $S^\mathrm{reg}\to0$, and the kinetic energy per site is given
as
\begin{equation}
  \varepsilon_\mathrm{kin} 
  = \frac{1}{N}\sum_{|k|<k_\mathrm{F}} (-2t\cos k)
  = -\frac{1}{\pi} 2t \sin k_\mathrm{F}
  \,.
\end{equation}
Since $z_k=1$, equation~(\ref{eq:sumrulegen2}) is fulfilled.

In the SC limit $g^2\gg1$, the leading term of $\mathcal{D}$ is
\begin{equation}\label{eq:app:DSC}
  \mathcal{D}
  \approx
  \tilde{W}e^2a\sin k_\mathrm{F}
  \,,
\end{equation}
having the form of the Drude term of free carriers with half-bandwidth
$\tilde{W}$.

Comparing this estimate of $\mathcal{D}$ to the integrated regular part of
the conductivity $S^\mathrm{reg}$, we find that the Drude part is negligible
in the SC limit, so that the kinetic energy is determined by
$S^\mathrm{reg}$ according to the f-sum rule~(\ref{eq:sumrulegen2}). To
demonstrate this, we shall not use equation~(\ref{eq:oc_sc}), but instead
start with $[\Re\,\sigma(\om)]_A$ following from
equations~(\ref{eq:Ae<}),(\ref{eq:Ae>}), (\ref{eq:abcddef}) and carry out
the integration over $\om$ before integrating over $k$. The dominating part
of $[\Re\,\sigma(\om)]_A$ for estimating $S^\mathrm{reg}$ in the SC limit
(corresponding to the contributions (A3) in figure~\ref{fig:trans_ic}) reads
\begin{eqnarray}\nonumber\fl\label{eq:app:resigmaSC}
  \Re\,\sigma(\om)
  =
  \frac{2\sigma_0}{\pi^2\om}
  \rme^{-2g^2}
  \sum_{s\geq1}\frac{(2g^2)^s}{s!}
  \int_0^\pi\rmd k' z_{k'} \theta(-E_{k'}-\eta)
  \\
  \times
  \int_0^\pi\rmd k'' z_{k''}\theta(E_{k''}+\eta)\delta(\om-s\om_0-E_{k''}+E_{k'})
  \,,
\end{eqnarray}
where we assumed $|E_k|<\om_0$ for all $E_k$ in the SC limit. The integration
over $\om$ gives
\begin{eqnarray}\label{eq:app:sregx}\nonumber\fl
  S^\mathrm{reg}
  =
  \frac{2\sigma_0}{\pi^2}\rme^{-2g^2}\sum_{s\geq1}\frac{(2g^2)^s}{s!}
  \int_0^\pi\rmd k' z_{k'} \theta(-E_{k'}-\eta)\\
  \times
  \int_0^\pi\rmd k'' z_{k''}\theta(E_{k''}+\eta)\frac{1}{s\om_0-E_{k''}+E_{k'}}
  \,.
\end{eqnarray}
To estimate $S^\mathrm{reg}$, we set $z_{k'},z_{k''}=1$ and assume
$E_{k'},E_{k''}\ll s\om_0$. Accordingly, the rational function in
equation~(\ref{eq:app:sregx}) is replaced by $1/s\om_0$ and the integrals over
$k',k''$ simply give $k_\mathrm{F}$ and $\pi-k_\mathrm{F}$,
respectively. The Fermi wavevector corresponding to the band population at
$T=0$ is $k_\mathrm{F}=n\pi$. The resulting estimate following from
equation~(\ref{eq:app:sregx}) is thus given as
\begin{equation}\label{eq:app:sreg}
  S^\mathrm{reg}
  =
  \frac{2\sigma_0}{\om_0}n(1-n)\sum_{s\geq1}\rme^{-2g^2}\frac{(2g^2)^s}{s!}\frac{1}{s}
  =
  \frac{2\sigma_0}{\om_0}n(1-n)\left<\frac{1}{s}\right>_{2g^2}
  \,.
\end{equation}

The brackets symbolise averaging with respect to the Poisson distribution
with the parameter $2g^2$, and the concentration dependence of
equation~(\ref{eq:app:sreg}) is in agreement with standard small-polaron
theory based on the atomic-limit approximation \cite{Ma90}.

Comparison of equation~(\ref{eq:app:sreg}) and equation~(\ref{eq:app:DSC})
reveals that the Drude term in the f-sum
rule~(\ref{eq:sumrulegen2}) is negligible in the SC limit, so that
$\varepsilon_\mathrm{kin}$ is proportional to $S^\mathrm{reg}$,
\begin{equation}\label{eq:app:ekin_sreg}  
  \varepsilon_\mathrm{kin}
  =
  -\frac{(2t)^2}{\om_0}n(1-n)\left<\frac{1}{s}\right>_{2g^2}
  \,.
\end{equation}

To make another check of the latter result, we compare
equation~(\ref{eq:app:ekin_sreg}) to the kinetic energy of one polaron at the
bottom of the band, obtained from equation~(\ref{eq:sc_band}) as
$t[\partial_k E_k]_{k=0}$ in \cite{FeLoWe00}. Keeping only the leading term
with respect to $\rme^{-g^2}$, setting $z_k=1$ and neglecting $E_k$ in
comparison to $\om_0$ as above, we obtain
\begin{equation}
  E_\mathrm{kin}
  =
  -\frac{(2t)^2}{\om_0}\left<\frac{1}{s}\right>_{2g^2}
  \,.
\end{equation}
Assuming $n\ll1$, $\varepsilon_\mathrm{kin}=n E_\mathrm{kin}$, in accordance
with equation~(\ref{eq:app:ekin_sreg}).


\section*{References}


\begin{thebibliography}{10}

\bibitem{CaDoLuPaMaGiRuChSa97}
  P Calvani, P Dore, S Lupi, A Paolone, P Maselli, P Giura, B Ruczicka, {S-W}
  Cheong, and W Sadowski, J. Supercond. {\bf 10},  293  (1997);
\item[]
  K~A M\"uller, J. Supercond. {\bf 12},  3  (1999).

\bibitem{SaAlLi95}
E~K~H Salje, A~S Alexandrov, and W~Y Liang, {\em Polarons and Bipolarons in
  High Temperature Superconductors and Related Materials} (Cambridge Univ.
  Press, Cambridge, 1995).

\bibitem{HoEd01}
M Hohenadler and D~M Edwards, J. Phys.: Condens. Matter {\bf 14},  2547
  (2002).

\bibitem{David_AiP}
D~M Edwards, Adv. Phys. {\bf 51},  1259  (2002).

\bibitem{HaMaLoKo04}
C Hartinger, F Mayr, A Loidl, and T Kopp, Phys. Rev. B {\bf 73},  024408
  (2006).


\bibitem{CaGrSt99}
M Capone, M Grilli, and W Stephan, Eur. Phys. J. B {\bf 11},  551  (1999).

\bibitem{HoNevdLWeLoFe04}
M Hohenadler, D Neuber, W {von der Linden}, G Wellein, J Loos, and H Fehske,
  Phys. Rev. B {\bf 71},  245111  (2005).

\bibitem{HoHaWeFe06}
M Hohenadler, G Hager, G Wellein, and H Fehske, J. Phys.: Condens. Matter
{\bf 19}, 255202 (2007).

\bibitem{Ho59a}
T Holstein, Ann. Phys. (N.Y.) {\bf 8},  325; {\bf 8}, 343  (1959).

\bibitem{Fr54}
H Fr\"{o}hlich, Adv. Phys. {\bf 3},  325  (1954).

\bibitem{ScWeWeAlFe05}
G Schubert, G Wellein, A~Wei{\ss}e~A Alvermann, and H Fehske, Phys. Rev. B {\bf
  72},  104304  (2005).

\bibitem{PhysRevB.44.7127}
Y.~Y. Suzuki, P. Pincus, and A.~J. Heeger, Phys. Rev. B {\bf 44},  7127
  (1991).

\bibitem{FeIhLoTrBu94}
H Fehske, D Ihle, J Loos, U Trapper, and H B\"{u}ttner, Z. Phys. B {\bf 94},
  91  (1994).

\bibitem{PeFiCaIa01}
C~A Perroni, G {De Filippis}, V Cataudella, and G Iadonisi, Phys. Rev. B {\bf
  64},  144302  (2001).

\bibitem{FeWeHaWeBi03}
H Fehske, G Wellein, G Hager, A Wei{\ss}e, and A~R Bishop, Phys. Rev. B {\bf
  69},  165115  (2004).

\bibitem{WeBiHoScFe05}
G Wellein, A~R Bishop, M Hohenadler, G Schubert, and H Fehske, Physica B {\bf
  378-380},  281  (2006).

\bibitem{LoHoFe06}
J Loos, M Hohenadler, and H Fehske, J. Phys.: Condens. Matter {\bf 18},  2453
  (2006).

\bibitem{LoHoAlFe06}
J Loos, M Hohenadler, A Alvermann, and H Fehske, J. Phys.: Condens. Matter {\bf
  18},  7299  (2006).

\bibitem{TeDe01}
J Tempere and J~T Devreese, Phys. Rev. B {\bf 64},  104504  (2001).

\bibitem{GuLaFi62}
V~L Gurevich, I~E Lang, and Yu~A Firsov, Fiz. Tverd. Tela {\bf 4},  1252
  (1962), [Sov. Phys.-Solid State 4, 918 (1962)].

\bibitem{AlMo95}
A~S Alexandrov and N Mott, {\em Polarons \& Bipolarons} (World Scientific,
  Singapore, 1995).

\bibitem{FeAlHoWe06}
H Fehske, A Alvermann, M Hohenadler, and G Wellein,  in {\em Polarons in Bulk
  Materials and Systems with Reduced Dimensionality}, {\em Proc. Int. School of
  Physics ``Enrico Fermi'', Course CLXI}, edited by G Iadonisi, J Ranninger,
  and G {De Filippis} (IOS Press, Amsterdam, Oxford, Tokio, Washington DC,
  2006), pp.\ 285--296.

\bibitem{SyHuBeWeFe04}
S Sykora, A H{\"u}bsch, K~W Becker, G Wellein, and H Fehske, Phys. Rev. B {\bf
  71},  045112  (2005).

\bibitem{CrSaCa05}
C~E Creffield, G Sangiovanni, and M Capone, Eur. Phys. J. B {\bf 44},  175
  (2005).

\bibitem{HoWeBiAlFe06}
M Hohenadler, G Wellein, A~R Bishop, A Alvermann, and H Fehske, Phys. Rev. B
  {\bf 73},  245120  (2006).

\bibitem{SyHuBe05}
S Sykora, A H\"{u}bsch, and K~W Becker, Eur. Phys. J B {\bf 51},  181  (2006).

\bibitem{Gr01}
A~C~M Green, Phys. Rev. B {\bf 63},  205110  (2001).

\bibitem{LangFirsov}
I~G Lang and Y~A Firsov, Zh. Eksp. Teor. Fiz. {\bf 43},  1843  (1962), [Sov.
  Phys. JETP {\bf 16}, 1301 (1962)].

\bibitem{Zubarev74}
D~N Zubarev, {\em Nonequilibrium statistical thermodynamics} (Plenum Press, New
  York, 1974).

\bibitem{Rick84}
G Rickayzen, {\em Green's Functions and Condensed Matter} (Academic Press,
  Inc., London, 1984).

\bibitem{pruschke_review}
T Pruschke, M Jarrell, and J~K Freericks, Adv. Phys. {\bf 44},  187  (1995).

\bibitem{Ma77}
P~F Maldague, Phys. Rev. B {\bf 16},  2437  (1977).

\bibitem{BaGrRi87}
D Baeriswyl, C Gros, and T~M Rice, Phys. Rev. B {\bf 35},  8391  (1987).

\bibitem{SePePl02}
D S\'{e}n\'{e}chal, D Perez, and D Plouffe, Phys. Rev. B {\bf 66},  075129
  (2002).

\bibitem{PhysRev.133.A171}
W Kohn, Phys. Rev. {\bf 133},  A171  (1964).

\bibitem{BoTrBa99}
J Bon\v{c}a, S~A Trugman, and I Batistic, Phys. Rev. B {\bf 60},  1633  (1999).

\bibitem{WeWeAlScFe05}
A Wei{\ss}e, G Wellein, A Alvermann, and H Fehske, Rev. Mod. Phys. {\bf 78},
  275  (2006).

\bibitem{FeTr2006}
H Fehske and S~A Trugman,  in {\em Polarons in Advanced Materials}, edited by
  A~S Alexandrov (Canopus Publishing and Springer Verlag GmbH, Bath (UK),
  2007).

\bibitem{Em93}
D Emin, Phys. Rev. B {\bf 48},  13691  (1993).

\bibitem{Ma90}
G~D Mahan, {\em Many-Particle Physics}, 2nd  ed. (Plenum Press, New York,
  1990).

\bibitem{AlKaRa94}
A~S Alexandrov, V~V Kabanov, and D~K Ray, Phys. Rev. B {\bf 49},  9915  (1994).

\bibitem{Ea84}
D~M Eagles, J. Phys. C: Solid State Phys. {\bf 17},  637  (1984).

\bibitem{LoSt89}
J Loos and J Straka, Czech J. Phys. {\bf 93},  316  (1989).

\bibitem{PhysRevLett.81.1517}
K.~H. Kim, J.~H. Jung, and T.~W. Noh, Phys. Rev. Lett. {\bf 81},  1517  (1998).

\bibitem{De98}
D~S Dessau, T Saitoh, C-H {Park}, Z-X {Shen}, P Villella, N Hamada, Y Moritomo,
  and Y Tokura, Phys. Rev. Lett. {\bf 81},  192  (1998).

\bibitem{AlBr99_2}
A~S Alexandrov and A~M Bratkovsky, J. Phys.: Condens. Matter {\bf 11},  L531
  (1999).

\bibitem{MiNaPrSaSv03}
A~S Mishchenko, N Nagaosa, N~V Prokof'ev, A Sakamoto, and B~V Svistunov, Phys.
  Rev. Lett. {\bf 91},  236401  (2003).

\bibitem{Ea84_2}
D~M Eagles, J. Phys. C: Solid State Phys. {\bf 17},  655  (1984).

\bibitem{FeLoWe00}
H Fehske, J Loos, and G Wellein, Phys. Rev. B {\bf 61},  8016  (2000).

\end{thebibliography}

\end{document}